\documentclass[lettersize,journal]{IEEEtran}
\usepackage{amsmath,amsfonts}
\usepackage{array}
\usepackage{textcomp}
\usepackage{multirow}
\usepackage{stfloats}
\usepackage{siunitx}
\usepackage{amssymb}
\usepackage{url}
\usepackage{verbatim}
\usepackage{graphicx}
\usepackage{caption}
\usepackage{algorithmic}
\usepackage{cite}
\usepackage{float}
\usepackage{color}
\usepackage{booktabs}
\usepackage{subcaption}
\usepackage[linesnumbered,ruled,lined]{algorithm2e}
\usepackage{balance}
\usepackage{multirow}

\usepackage{tikz}
\usepackage{lipsum}
\usepackage[english]{babel}
\usepackage{enumitem}
\usepackage{subcaption}
\usepackage[bottom]{footmisc}

\usepackage{amsthm}

\makeatother

\newlength{\RoundedBoxWidth}
\newsavebox{\GrayRoundedBox}
   {\setlength{\RoundedBoxWidth}{\dimexpr#1}
    \begin{lrbox}{\GrayRoundedBox}
       \begin{minipage}{\RoundedBoxWidth}}%
   {   \end{minipage}
    \end{lrbox}
    \begin{center}
    \begin{tikzpicture}%
       \draw node[draw=black,fill=black!10,rounded corners,%
             inner sep=2ex,text width=\RoundedBoxWidth]%
             {\usebox{\GrayRoundedBox}};
    \end{tikzpicture}
    \end{center}}

\newcommand{\etal}{\textit{et al.}}

\begin{document}

\title{User-Generated Content and Editors in Games: A Comprehensive Survey}

\author{Yuyue Liu, Haihan Duan,~\IEEEmembership{Member,~IEEE}, Wei Cai,~\IEEEmembership{Senior Member,~IEEE}% <-this % stops a space
%
% \thanks{Manuscript received 29 February 2024; firstly revised 29 April 2024; secondly revised 30 May 2024; accepted 11 June 2024.}
\thanks{This work was supported by Guangdong-Hong Kong-Macao Joint Laboratory for Emotional Intelligence and Pervasive Computing, Artificial Intelligence Research Institute, Shenzhen MSU-BIT University. (Corresponding Author: Haihan Duan)}
\thanks{Yuyue Liu is with Guangdong-Hong Kong-Macao Joint Laboratory for Emotion Intelligence and Pervasive Computing, Artificial Intelligence Research Institute, Shenzhen MSU-BIT University (as a visiting student), and School of Data Science, The Chinese University of Hong Kong, Shenzhen, China. E-mail: yuyueliu@link.cuhk.edu.cn.}
\thanks{Haihan Duan (corresponding author) is with Guangdong-Hong Kong-Macao Joint Laboratory for Emotion Intelligence and Pervasive Computing, Artificial Intelligence Research Institute, Shenzhen MSU-BIT University. E-mail: duanhaihan@smbu.edu.cn.}
\thanks{Wei Cai is with School of Engineering and Technology, University of Washington, Tacoma, WA, USA. E-mail: weicaics@uw.edu.}
}

\markboth{IEEE Transactions on Games,~Vol.~XX, No.~X, July~2024}%
{Shell \MakeLowercase{\textit{et al.}}: A Sample Article Using IEEEtran.cls for IEEE Journals}


\maketitle

\begin{abstract}
User-Generated Content (UGC) refers to any form of content, such as posts and images, created by users rather than by professionals. In recent years, UGC has become an essential part of the evolving video game industry, influencing both game culture and community dynamics. The ability for users to actively contribute to the games they engage with has shifted the landscape of gaming from a one-directional entertainment experience into a collaborative, user-driven ecosystem. Therefore, this growing trend highlights the urgent need for summarizing the current UGC development in game industry. Our conference paper has systematically classified the existing UGC in games and the UGC editors separately into four types. However, the previous survey lacks the depth and precision necessary to capture the wide-ranging and increasingly complex nature of UGC. To this end, as an extension of previous work, this paper presents a refined and expanded classification of UGC and UGC editors within video games, offering a more robust and comprehensive framework with representative cases that better reflects the diversity and nuances of contemporary user-generated contributions. Moreover, we provide our insights on the future of UGC, involving game culture, game genre and user creative tendencies, artificial intelligence, its potential ethical considerations, and relationship between games, users and communities.
\end{abstract}


\section{Introduction} \label{sec_intro}

User-Generated Content (UGC) refers to any form of content---such as text, posts, images, videos, reviews, and other media—created by users or consumers, rather than by official developers of online platforms \cite{min2019blockchain,naab2017studies}. In recent years, UGC has become a cornerstone of the modern video game industry, providing players with the tools to create, share, and engage in creative content beyond the confines of traditional game development. This dynamic between players and games not only enhances engagement but also drives the cultural and economic growth of gaming communities \cite{zhao2009survey,sun2017motivation}. Given its increasing importance in ensuring the long-term sustainability of video games, UGC is a critical area for ongoing research.

The scope of UGC in video games is vast, encompassing everything from user-designed character appearances to modifications (MODs) that alter game code. The most notable examples in the history of video game development are \textit{Counter-Strike (CS)}\footnote{https://blog.counter-strike.net/} and \textit{Defense of the Ancients (DOTA)}\footnote{https://www.dota2.com/home}, where \textit{Counter-strike} originated from a MOD of \textit{Half-Life}\footnote{https://www.half-life.com/}, a first-person shooting (FPS) video game, and \textit{DOTA} was a multiplayer online battle arena (MOBA) MOD of \textit{Warcraft III: Reign of Chaos}\footnote{https://playwarcraft3.com/}. UGC enhances players' gaming experiences by promoting greater engagement and giving them a sense of control. Through such user-centered content creations, users can modify game mechanics, such as altering character models or changing gameplay rules with MODs, leading to innovative experiences that increase enjoyment and satisfaction \cite{kasapakis2017user, diaz2021building}.

In recent years, research on UGC has primarily examined its forms and impact across various specific fields, including online brand communities \cite{estrella2017different}, brands marketing \cite{christodoulides2012memo}, tourism \cite{ukpabi2018drives}, hotel experiences \cite{barreda2013analysis}, digital well-being \cite{cuomo2020user}, social media \cite{luca2015user}, etc., with limited attention given to a systematic analysis of UGC in video games \cite{lastowka2007user,lundmark2015effects, brunt2020influence}. For instance, in 2017, Estrella \etal \cite{estrella2017different} categorized the impact of UGC on video game sales and distribution, highlighting its potential role in the commercial success of games. However, their taxonomy lacks a detailed breakdown of in-game UGC, such as MOD that alter game rules, mechanics, or assets---components that are pivotal to understanding UGC's broader influence within the gaming ecosystem. A systematic review of these types of UGC, particularly with a focus on how they modify gameplay mechanics or alter in-game resources, remains largely unexplored in academic literature. Therefore, addressing this gap is crucial for understanding the full scope of UGC's influence on both game development and player engagement, as well as its potential to reshape the future of game design and community involvement.

As outlined in our conference paper \cite{duan2022user}, we seek to fill this academic gap by providing systematic classifications of existing UGC and editors within the video game industry. The major contributions can be concluded as follows:

\begin{itemize}
    \item  We construct a decision tree-like classification Venn diagram based on the influence of UGC on video games, divided UGC and UGC editors into four categories, and listed the corresponding typical cases. 
    \item  We enumerate the propagation methods of UGC in video games, such as player community sharing, circulation in trading markets, and combination with original games, which boost a positive circulation of UGC and improve user's engagement.
    \item  We propose our visions for the future development of UGC in the metaverse, emphasizing its potential to drive immersive experiences, foster collaborative virtual economies based on blockchain, and reshape social interactions within expansive and player-driven virtual worlds.
\end{itemize}

However, the classification framework suggested in the conference version \cite{duan2022user} has notable limitations, since it does not fully capture the diversity of UGC, and the distinctions between certain categories are slightly ambiguous. Therefore, compared with the conference version \cite{duan2022user}, this work mainly encompasses the extension of the following aspects:

\begin{itemize}
    \item This version prioritizes ``user experience" as the central criterion, refining the previous taxonomy and introducing a more detailed parallel classification, with representative games for each category. In this research, UGC is classified into two primary categories based on its impact on the gaming experience: the changes of gameplay and resource file, as gameplay is related to the ``expansion" of gaming experience, and resource files affects the ``improvement" part. 
    \item This version differentiates UGC editors from general game editors, offering unique insights into UGC editors by defining two distinct content generations: the professional generation, for skilled developers creating indie games, and the ordinary user generation, for users without technical expertise. The latter focuses on simpler, fan-made content, commonly known as UGC, enriching games without requiring advanced development skills.
    \item This version presents an in-depth analysis of UGC across various dimensions, examining its evolution beyond video games, involving game culture, game genre and user creative tendencies, artificial intelligence, its potential ethical considerations. It also explores the intricate relationship between users, games, and and relationship between games, users and communities, highlighting how UGC shapes interactions and envisions future collaborative dynamics that foster deeper engagement and development.
\end{itemize}

The following Section \ref{sec_rw} will offer a brief overview of the previous research. Then, we will separately elaborate the refined classification and representative cases of UGC and editors in Section \ref{sec_ugc} and \ref{sec_editor}. In Section \ref{sec_dis}, we explore the future of video game UGC and its potential challenges, while Section \ref{sec_conclusion} provides a conclusion.

\section{Related Work} \label{sec_rw}

Research into UGC in video games has expanded significantly over the years, reflecting the growing importance of player-created content in shaping game experiences, environments and communities. Early studies on UGC primarily focused on its role in user motivation and extending the lifecycle of video games. Lastowka \etal \cite{lastowka2007user} explored the legal implications of UGC creation within virtual worlds. Daugherty \etal \cite{daugherty2008exploring} studied the motivations driving users to create UGC. Kasapakis \etal \cite{kasapakis2017user} discussed best practices for leveraging UGC effectively in pervasive games using \textit{Barbarossa} as a case. Wirman \cite{wirman2009productivity} identified five dimensions of player productivity in the video game industry and emphasized the importance of other users and communities to player productivity. Lindberg \etal \cite{lundmark2015effects} conducted a quantitative study involving surveys of members from four online video game UGC communities, which examined the influence of intrinsic motivation, extrinsic motivation, and the availability of toolkits in encouraging players to participate in the creation of UGC for video games. Researchers like Sotamaa \cite{sotamaa2015online} have examined the participatory nature of UGC, where players act as co-creators, enhancing the collaborative nature of game development. Moreover, some studies have examined the presentations of UGC in video games. For example, Postigo \cite{postigo2008video} analyzed user demands and developer-user conflicts through two prominent MODs: \textit{Duke It Out in Quake} for \textit{Quake III}\footnote{http://www.q3arena.com/} and \textit{GI Joe} for \textit{Battlefield 1942}\footnote{https://en.wikipedia.org/wiki/Battlefield\_1942}. However, the existing studies mainly focus on specific game types or individual cases, while, besides our conference paper \cite{duan2022user}, there currently lacks a universal framework with comprehensive and systematic classification of UGC, which can be applied to most video games.

On the other hand, development toolkits and editors are essential for the creation of UGC, as they directly influence the user's creation experience. Several researchers have explored how these game editors affect UGC creation, leveraging various technologies. Liapis \etal \cite{liapis2013sentient} introduced the Sentient Sketchbook, a tool that assists designers in creating game levels by using map sketches to reduce the workload of level design. Ng \etal \cite{ng2018situated} focused on the augmented reality (AR) environment, investigating the design and concept of level editors specifically for AR games. Currently, some games have begun experimenting with new co-creation models, granting players' creative rights to unlock their potential and extend the game's lifespan. In the context of co-creation in video games, Davidovici-Nora \cite{davidovici2009dynamics} highlighted a hybrid innovation model implemented in \textit{World of Warcraft (WoW)}\footnote{https://worldofwarcraft.blizzard.com}, a popular massively multiplayer online role-playing game (MMORPG). In addition, the integration of Artificial Intelligence (AI) in user content generation has gained significant attention from the academia, since AI can significantly lower the barriers to user creation and reduce production costs. For instance, Summerville \etal \cite{summerville2016super} examined the use of Long Short-Term Memory (LSTM) recurrent neural networks to generate levels for \textit{Super Mario Bros}. Guzdial \etal \cite{guzdial2016game} introduced an unsupervised process for generating full game levels based on gameplay video. Awiszus \etal \cite{awiszus2021world} introduced ``World-GAN", a generative model for \textit{Minecraft}\footnote{https://www.minecraft.net} worlds using natural language processing (NLP) technology to create larger, more complex worlds with specified terrain. Furthermore, Guzdial \etal \cite{guzdial2019friend} discussed in detail how the design of AI-driven level editors influences creators' experiences. However, previous researches have primarily focused on innovations in the design and implementation of tools and editors for generating UGC, without clearly distinguishing between UGC editors and other general game editors. Besides, the editors mentioned above remain largely inaccessible to users, with their primary target users still being professional game level designers. In summary, there is a lack of systematic classification and comprehensive reviews of the existing UGC editors.

\begin{figure*}
    \centering
    \includegraphics[width=2.0\columnwidth]{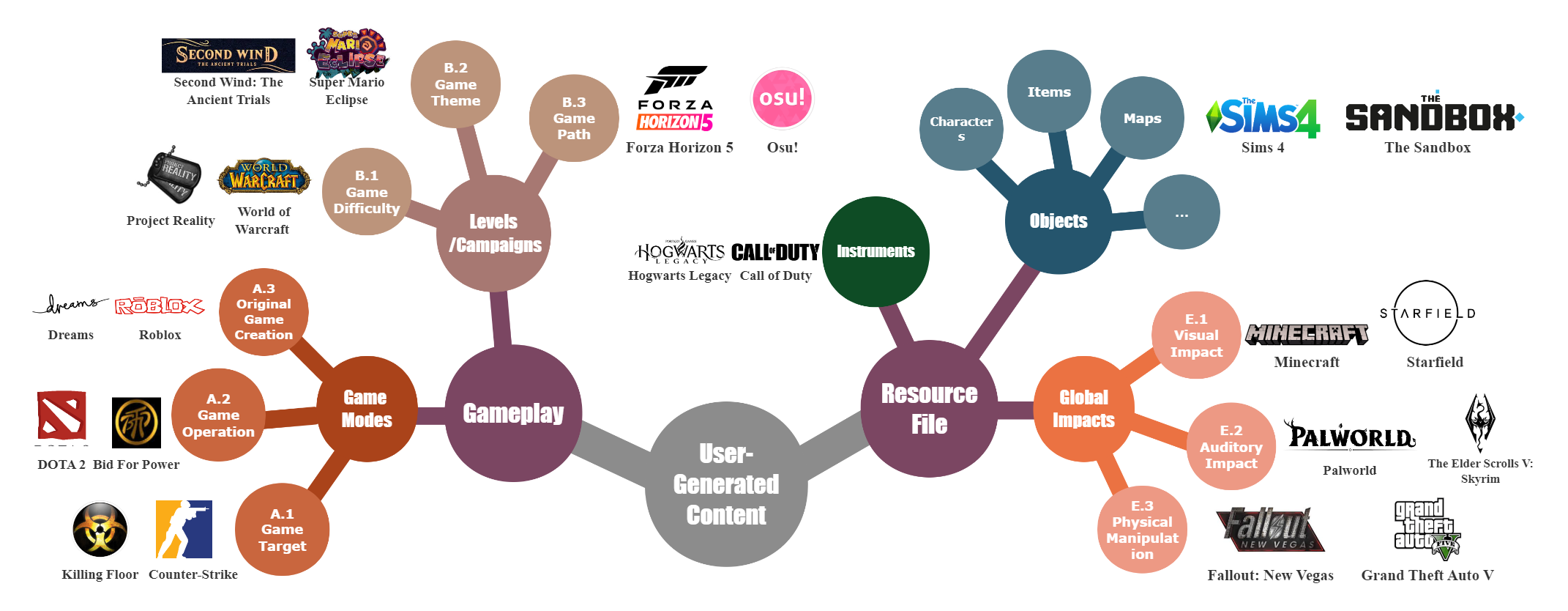}
    \caption{Parallel classification diagram of user-generated content and corresponding representative games for each type}
    % \caption{Parallel classification diagram of user-generated content and corresponding representative games for each type. Since there are many types of game objects, only a few representative examples are selected for analysis in this article (e.g., \textit{Sims 4} and \textit{The Sandbox}).}
    \label{fig_ugc}
\end{figure*}

\section{Categories of User-Generated Content} \label{sec_ugc}

Our conference version \cite{duan2022user} has taken a decision tree-style classification, but we also realized that UGC lends itself more naturally to a parallel classification approach. According to J. R. Quinlan's research \cite{quinlan1986induction}, dividing UGC into rigid categories is insufficient since some modifications might simultaneously change the default rules and game resources. For those which changed the resource files to create a new game mode, the original decision tree-style classification Venn diagram \cite{duan2022user} would have ignored changes to the game resource file, because the decision to determine whether it was a game mode change occurred in front of the resource file, and once the judgment was true, it would not be further judged. This classification treats UGC categories as mutually exclusive, when in reality, most UGC is a blend of multiple types. Given these overlaps, a more flexible classification framework is necessary. Furthermore, the original paper's four-category classification \cite{duan2022user} does not fully encapsulate the range of UGC in video games. Drawing from Postigo’s taxonomy of game fan-programmers \cite{postigo2007mods}, which divides those amateur game developers into three categories: ``modders" who change the game mode, ``mappers" who change the level, and ``skinners" who change the game character and other resources, we regard users as the key to UGC and split it into two basic categories according to the users' gaming experience including the change of gameplay and resource files.

\textbf{(1) The change of gameplay:} By modifying the default rules or adding new levels, users can expand their gaming experience. Adams \cite{adams2014fundamentals} provided a non-rigorous definition of gameplay, dividing it into two aspects: the ``challenge" presented to players and the ``actions" they need to take. Building on this work, we broaden the definitions of ``challenge" and ``action" to encompass all possible gameplay situations. Generally speaking, we classify the change of gameplay into two directions including game modes and game levels/campaigns. The game mode changes are more focused on the core rules and mechanics of the game, while the changes of game levels/campaigns tend to rely on existing modes to generate new challenges. Note that, according to the study of Adams \cite{adams2014fundamentals}, we consider that the emphasis of the change of gameplay should on the novelty of experience comparing to the change of resource file. For example, uploading a user-made map without any change of game framework or storylines should be considered as a resource file. 

\textbf{(2) The change of resource files:} Users can modify the game files or upload new resources to enhance their gaming experience. The most common MODs in the video game community typically involve resource files that do not alter gameplay or core mechanics. Instead, they achieve various purposes by adding or overwriting files in the game's installation folder. These modifications can range from practical improvements, like automating repetitive tasks, to more playful changes, such as replacing character models with real-life celebrities in RPGs. In this paper, we consider to classify these resource files into three categories: new objects, new tools, and global modifications.

As shown in Figure \ref{fig_ugc}, we construct a parallel classification bubble diagram with corresponding representative games for each type. Note that, the UGC we study is categorized from the perspective of the user, which does not mean that each game belongs to only one specific category, in other words, users may create different contents within the same game. Moreover, since there are many types of game objects, only a few representative examples are selected for analysis in this article (e.g., \textit{Sims 4} and \textit{The Sandbox}). In the following subsections, we will clarify the definition and further discuss the characteristics and prototypes in detail.

\subsection{User-generated Game Modes}\label{subsec_mode}

User-generated game modes introduce innovative gameplay experiences by modifying fundamental rules or objectives. In this paper, we assert that the emphasis on game modes should reflect the mechanics of the original games. As mentioned in the conference version \cite{duan2022user}, new game modes need not be entirely distinct from their base games; rather, they should provide novel gaming experiences that deviate from the original rules or targets. To better understand how these modifications manifest across different game types, we categorize them into three aspects: game targets, operations, and original game creation. Each of these categories will be elaborated upon in the following sections.

\subsubsection{A.1 The change of game target}

\begin{figure}[b!]
	\centering
	\begin{minipage}[t]{0.49\columnwidth}
		\centering
		\includegraphics[width=\columnwidth]{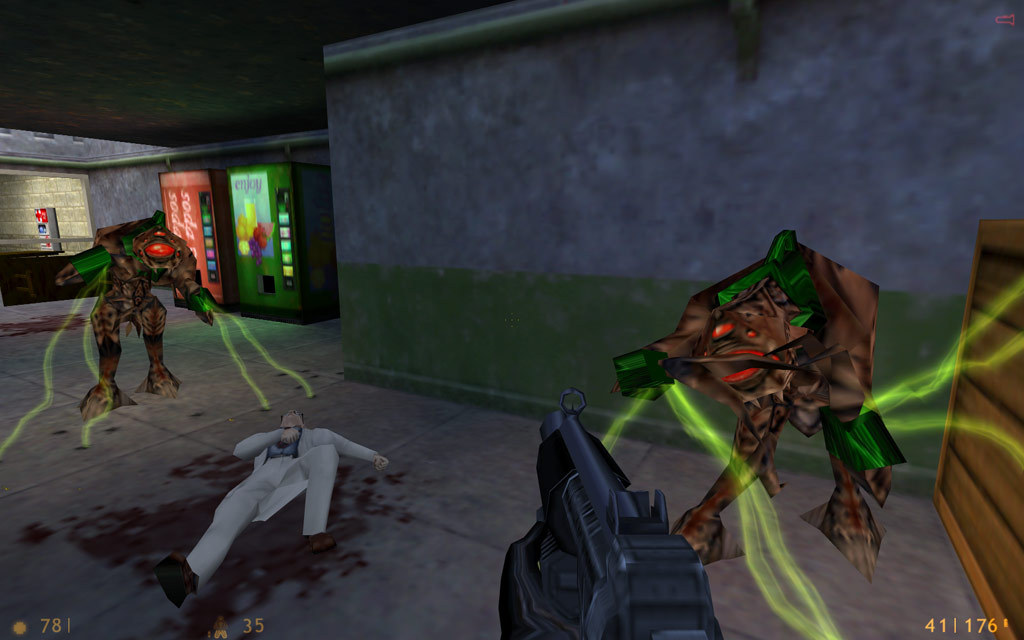}
		\subcaption{\textit{Half-Life}}
	\end{minipage}
	\begin{minipage}[t]{0.49\columnwidth}
		\centering
		\includegraphics[width=\columnwidth]{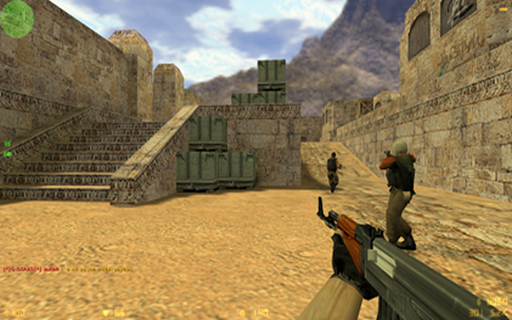}
		\subcaption{\textit{Counter-Strike}}
	\end{minipage}
        \begin{minipage}[t]{0.49\columnwidth}
		\centering
		\includegraphics[width=\columnwidth]{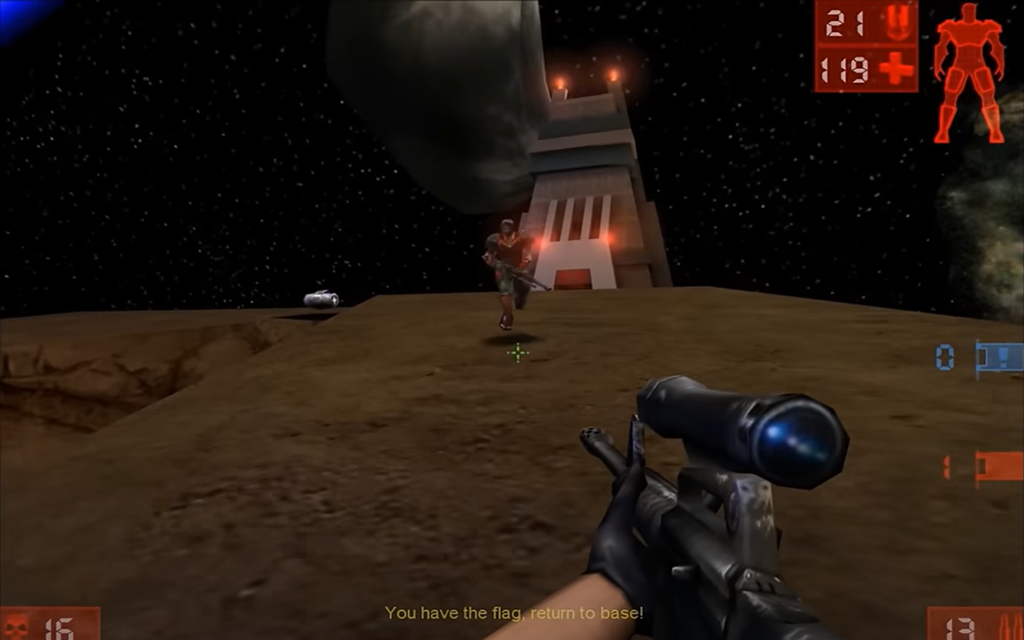}
		\subcaption{\textit{Unreal Tournament}}
	\end{minipage}
	\begin{minipage}[t]{0.49\columnwidth}
		\centering
		\includegraphics[width=\columnwidth]{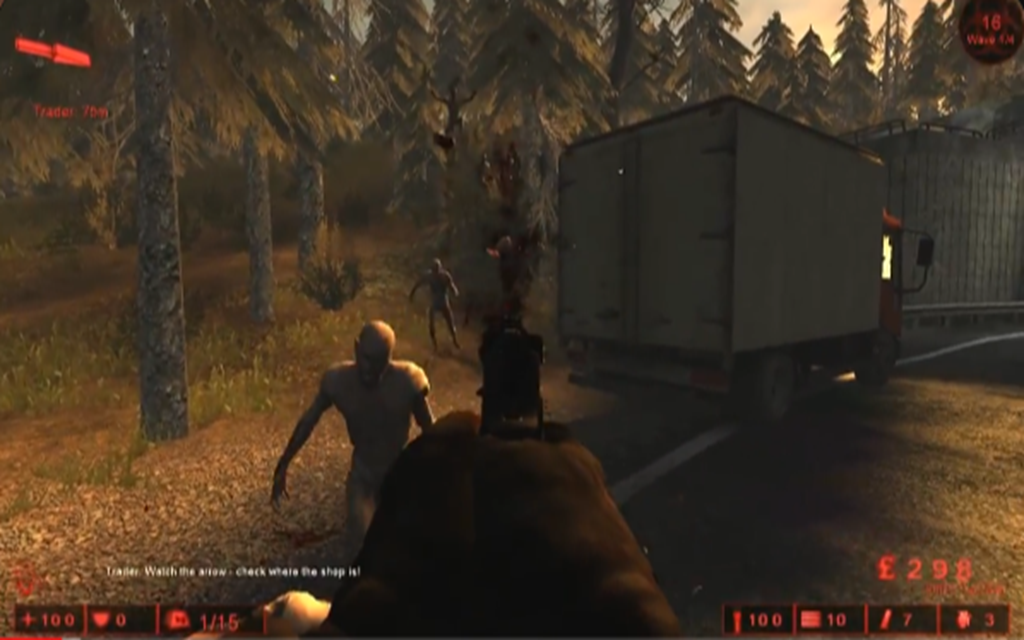}
		\subcaption{\textit{Killing Floor}}
	\end{minipage}
        \caption{Game target changes in first-person shooting games}
        \label{fig_game_target_change}
\end{figure}

The target of the game often directly determines the condition to win in most competitive games, i.e., a new game mode can be generated by changing some rules of the original game. The most notable case is the aforementioned \textit{Counter-strike} (Figure \ref{fig_game_target_change} (b)), which is originally developed by Minh ``Gooseman" Le and Jess Cliffe in 1999. As a modification of \textit{Half-Life} (Figure \ref{fig_game_target_change} (a)), it changed the gameplay of \textit{Half-Life}, which required users to solve puzzles and follow the storyline, instead it divided players into two teams---terrorists and the counter-terrorists---to fight against each other. Depending on the map, there are four types of scenarios \footnote{https://counterstrike.fandom.com/wiki/Map}: hostage rescue, bomb defusal, assassination and escape, where the official maps of the fourth mission type ``escape" has been removed before the game's release. The creators of the initial modification have obviously changed the game target of \textit{Half-Life}, abandoning chapters that rely on story support. Compared with the team deathmatch mode that players are fighting alone most of the time, \textit{Counter-strike} chooses to rely on the direct confrontation between two opposing teams to win. 

The change of game target is usually more intuitive and less connected to the original work. Therefore, we believe that changing the game target means changing some specific rules of the basic game, or even the storyline. In other word, it is more appropriate to be regarded as a standalone spin-off game. For another example, \textit{Killing Floor}\footnote{https://www.killingfloor2.com/} (Figure \ref{fig_game_target_change} (d)) was released in 2005 as a MOD for \textit{Unreal Tournament 2004 (UT)}\footnote{https://www.utzone.de/} (Figure \ref{fig_game_target_change} (c)) which had completely different game modes. \textit{UT} is an arena shooter focused on fast-paced, competitive multiplayer matches. The game target depends on different game mode but usually revolves around killing opponents to score points or capturing objectives to win the match. But \textit{Killing Floor} aims to survive multiple waves of horror-themed enemies and defeat the final boss in a cooperative setting. In \textit{Killing Floor}, the creators also changed the sci-fi theme from the original to a gruesome and bloody theme filled with zombie-like creatures, which could also be regarded as a change in the game theme, but here we consider the change of game target is dominant.

\subsubsection{A.2 The change of game operation}

\begin{figure}[b!]
	\centering
	\begin{minipage}[t]{0.49\columnwidth}
		\centering
		\includegraphics[width=\columnwidth]{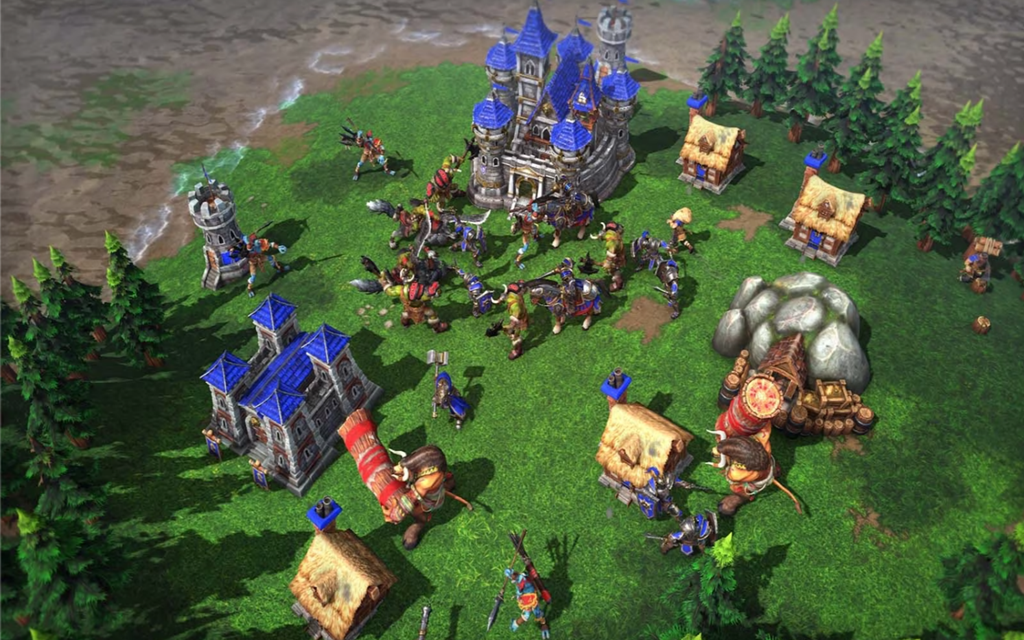}
		\subcaption{\textit{Warcraft III}}
	\end{minipage}
	\begin{minipage}[t]{0.49\columnwidth}
		\centering
		\includegraphics[width=\columnwidth]{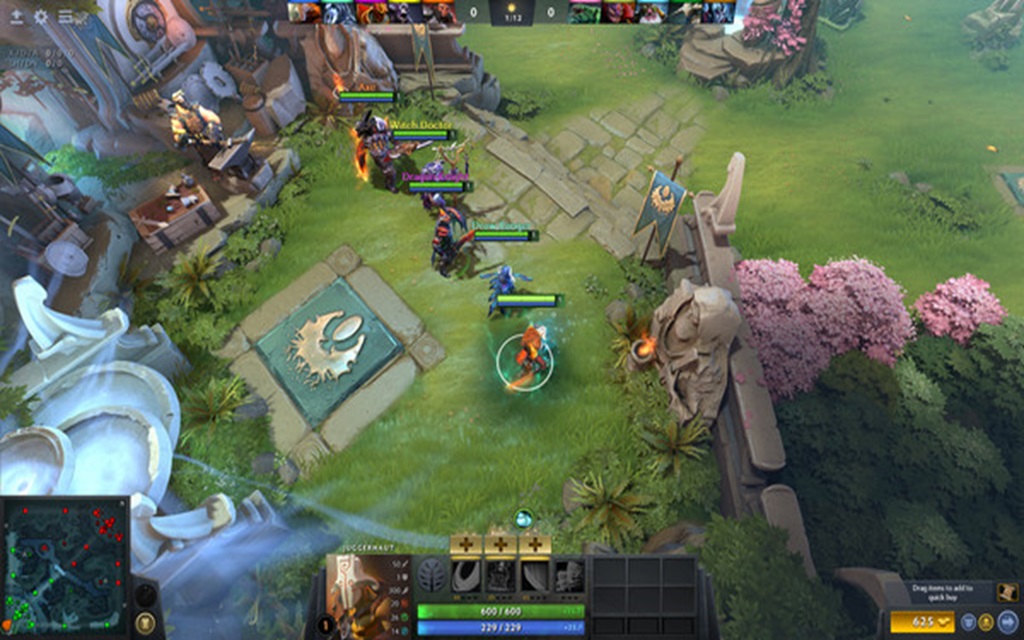}
		\subcaption{\textit{DOTA}}
	\end{minipage}
        \caption{Game operation changes in \textit{DOTA}}
        \label{fig_game_operation_change}
\end{figure}

Game operation refers to the mechanics of how the game is played, distinct from the game’s target or objective. While a modification changes the game’s target, the core gameplay mechanics, or operations, might remain consistent. For instance, \textit{Counter-strike} mentioned above has changed the rules of winning the game, but in terms of operation, it is still a first-person shooting game. However, changing the game operation often involves changing the game target. A simple case is altering the game's perspective, such as switching from first-person to third-person view, without changing other gameplay elements. If more modifications are added, the target and tempo of the game may also change. \textit{Bid For Power(BFP)}\footnote{https://bidforpower.net/} is a complete conversion for \textit{Quake III} with gameplay that is quite different from the original game. It changes the way the original first-person shooter works, allowing players to control the Ki-powered superhero from a third-person perspective, and it is faster paced than the original game. 

In the context of more complex modifications, we argue that the concepts discussed in this paper can be integrated. User-generated modifications can alter both game operations and targets, resulting in a comprehensive shift in gameplay dynamics. A representative case is \textit{DOTA} (Figure \ref{fig_game_operation_change} (b)). Originally, the \textit{Warcraft III: Reign of Chaos} (Figure \ref{fig_game_operation_change} (a)) allows players to command armies to fight against enemies' bases in a real-time strategy (RTS) game, while \textit{DOTA} changed the mode of controlling an army to only controlling a hero character. Comparing with \textit{Warcraft III}, \textit{DOTA} divides players into two teams of five for 45-90 minutes, with each player choosing the unique hero to fight. From the gameplay point of view, \textit{Warcraft III} is more open, and this feature is more prominent in the subsequent \textit{World of Warcraft (WOW)}. Although \textit{DOTA} and \textit{Warcraft III} share similarities in worldview and character setting, as Remi Smith \cite{Smith_Fodora_Guzik_2023} says, just like chess and Monopoly that both require strategy and experience, but they are fundamentally different.

\subsubsection{A.3 Original game creation} \label{subsec_original_game_creation}

\begin{figure}[h!]
	\centering
	\begin{minipage}[t]{0.49\columnwidth}
		\centering
		\includegraphics[width=\columnwidth]{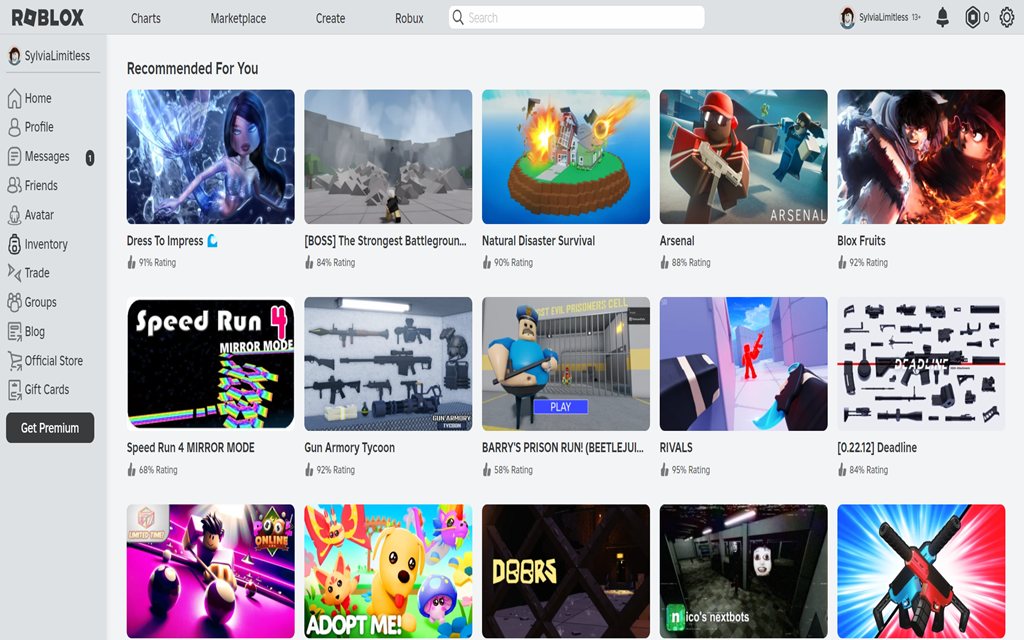}
		\subcaption{Game Community}
	\end{minipage}
	\begin{minipage}[t]{0.49\columnwidth}
		\centering
		\includegraphics[width=\columnwidth]{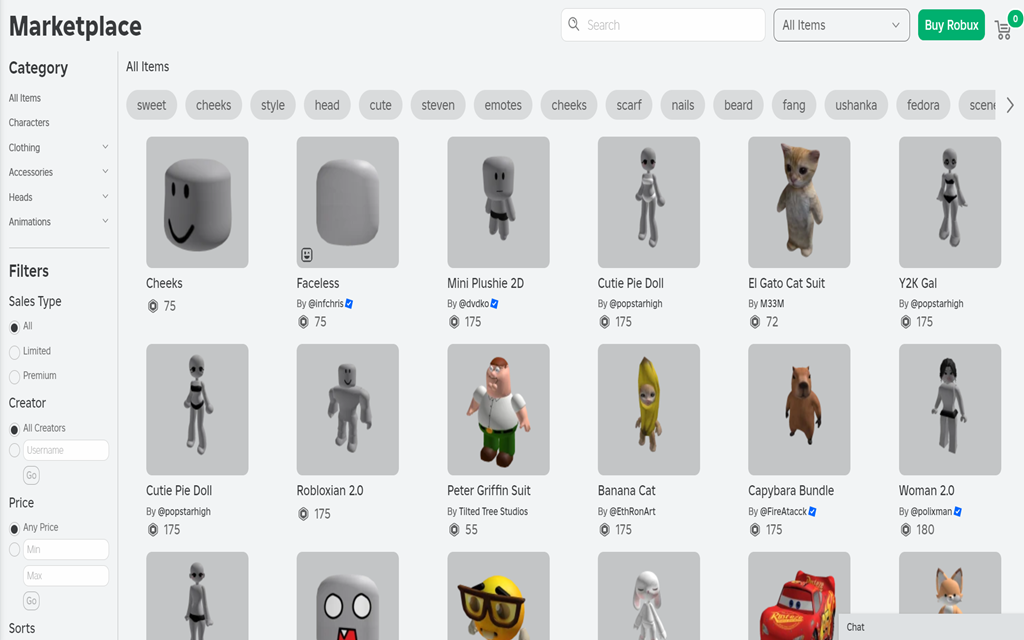}
		\subcaption{Marketplace}
	\end{minipage}
        \caption{\textit{Roblox} encourages multiple original game creations}
        \label{fig_roblox}
\end{figure}

This category differentiates itself from traditional game modifications, as illustrated by \textit{Roblox}\footnote{https://www.roblox.com/
}. These games only provide basic rules, environments, materials, and operations, and players can create entirely new game modes by altering both gameplay objectives and mechanics. These alterations include changes to both game targets and operations, while remaining closely aligned with the original framework. In this way, the original game creation transforms the original game into an open platform, functioning like a game editor or a resource library for users. \textit{Roblox} encourages users to generate their own game using internal resources and upload it on the community (Figure \ref{fig_roblox} (a)) so that other users can easily acquire what are they interested in and even pay for it through the marketplace (Figure \ref{fig_roblox} (b)). While some developers offer corresponding game editors, few platforms provide the same level of freedom as \textit{Roblox}. The use of \textit{Roblox Studio} allows players to create games with greater ease, effectively reducing barriers associated with modifying game code. As noted by Young-joo \etal \cite{KANG2024100697}, \textit{Roblox} blurs the distinction between game creators and players. Comparing with other games, this platform functions similarly to a marketplace, where players can buy and sell content, while developers are responsible for establishing rules and maintaining order within the ecosystem.

\textit{``Dreams"}\footnote{https://www.playstation.com/de-de/games/dreams/}, developed by \textit{Media Molecule}\footnote{https://www.mediamolecule.com/} for the \textit{Play Station 4}\footnote{https://www.playstation.com/de-de/ps4/}, serves as another notable example of original game creations. As an innovative and expansive game, it offers robust development tools that give players a high degree of freedom to create and share various types of content, including animations, games, and music. Users can design everything from simple mini-games to complex, fully-featured experiences, utilizing the platform's capabilities to sculpt 3D models, apply textures, create animations, compose music, and develop interactive elements through an intuitive interface. Both \textit{Roblox} and \textit{``Dreams"} illustrate the potential of user-generated content to transform traditional gaming experiences into open, collaborative platforms.

\subsection{User-generated Game Levels/Campaigns}

As another type of gameplay change, user-generated levels can be regarded as a continuation of the original game. Instead of changing the rules by modifying the game code, users can choose to create a whole new level which is the most fundamental difference between new mode and new level. Such changes do not usually result in an independent game, because users may generate these levels because they are still unsatisfied after completing the game and want to chase for some harder challenges. Depending on the purpose of creating new levels, we divide it into two directions: the change of game difficulty and the change of game theme. The following subsections will analyze their characteristics and discuss some typical examples respectively.

\subsubsection{B.1 The change of game difficulty} \label{subsubsec_game_difficulty}

In the conference version \cite{duan2022user}, we classified the change of game tempo as ``a change of game mode". However, further research revealed a lack of consensus on whether changes in tempo truly constitute gameplay changes. The term ``tempo" originates from music terminology, where it defines the pace or speed of a composition. Therefore, the closest examples to its original meaning are music games (sometimes also be called rhythm games) that users need to operate according to the rhythm of the music such as \textit{Muse Dash}\footnote{https://musedash.peropero.net}. In this paper, we expand the definition of game tempo beyond rhythm games to apply it to all types of games, acknowledging that tempo can take different forms depending on the game’s default rules and objectives. In card games, for instance, the number of cards in the deck and the time limit of each turn affect players' stresses in the fight. In MMORPGs such as \textit{World of Warcraft}, factors such as cooldown times or skill damage affect the time-to-kill (TTK), which alters the pace of combat, impacting gameplay difficulty and experience.

On the other hand, adjustments to the game’s tempo, such as increasing the number of enemies, do not significantly alter the original game and therefore cannot be classified as a new game mode. Instead, these adjustments are better understood as modifications in difficulty, which can be treated as a new user-generated level. But it is crucial to note that this definition is used only to distinguish it from a new game mode and does not actually generate a new level or campaign. For instance, most RPG games offer various difficulty levels that do not alter the narrative or core challenges but instead adjust elements like enemy health and damage. In such cases, the levels with higher difficulty simply make the game more challenging without introducing fundamentally new content.

\begin{figure}[h!]
	\centering
	\begin{minipage}[t]{0.49\columnwidth}
		\centering
		\includegraphics[width=\columnwidth]{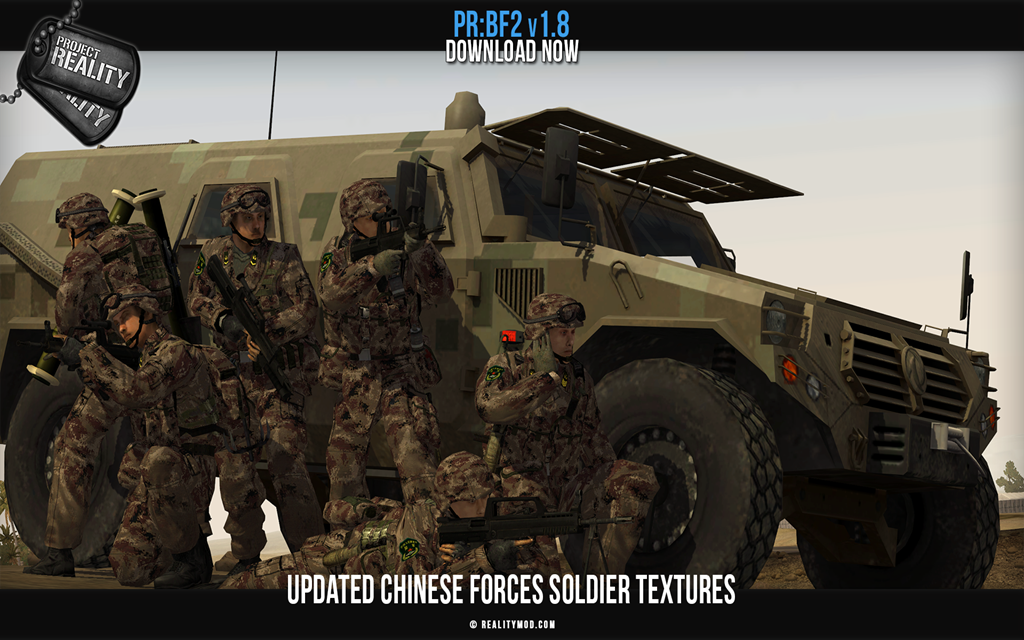}
		\subcaption{New Forces Solders}
	\end{minipage}
	\begin{minipage}[t]{0.49\columnwidth}
		\centering
		\includegraphics[width=\columnwidth]{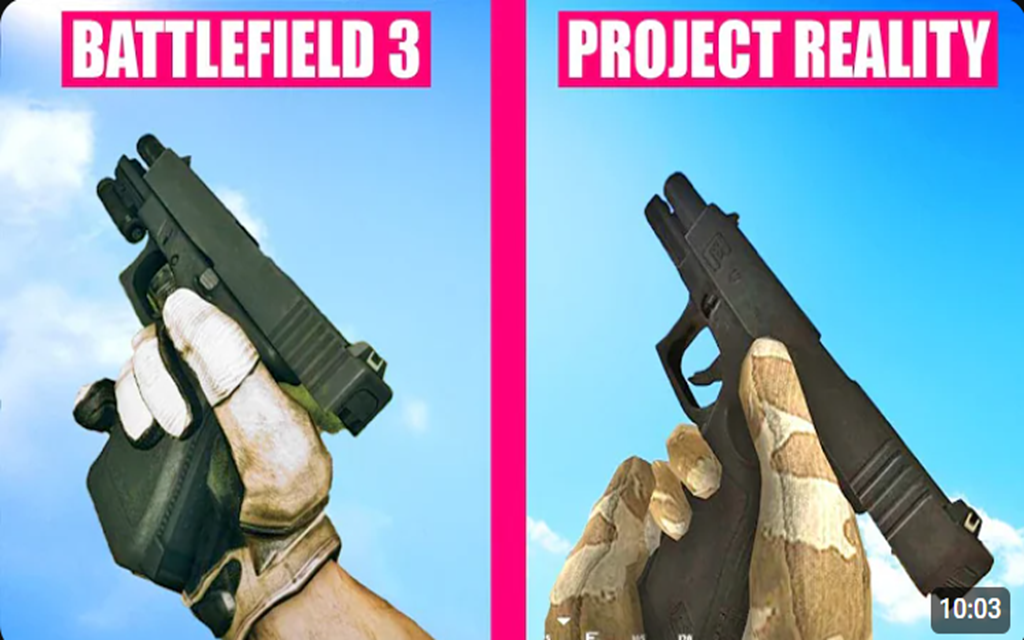}
		\subcaption{Weapon Comparison}
	\end{minipage}
	\caption{Game tempo changes in \textit{Project Reality}}
        \label{fig_project_reality}
\end{figure}

\textit{Project Reality (PR)}\footnote{https://www.realitymod.com/} is a single and multiplayer PC game based on the \textit{Battlefield 2}, which was developed by \textit{Digital Illusion of Sweden (DICE)}\footnote{https://www.dice.se/}. The production team modified the original game to create a more realistic gaming experience by improving the ballistic system and providing a shorten time to kill(TTK). Compared with \textit{Battlefield 2}\footnote{https://www.ea.com/games/battlefield}, \textit{Project Reality} introduces armed and irregular forces from around the world and enlarge the scale of battlegrounds, as shown in Figure \ref{fig_project_reality} (a). Developers recalculate the damage mechanics, taking more realistic factors into consideration such as the distance of bullet drop and weapon caliber damage. Meanwhile, it is worth noting that in \textit{Project Reality} players can quickly identify enemies or friendlies from the difference in the sounds of weapons, which is hard in \textit{Battlefield 3}\footnote{https://www.ea.com/de-de/games/battlefield/battlefield-3} because it allows everyone use any weapon and the sounds are pretty similar, as shown in Figure \ref{fig_project_reality} (b). Those modifications shape the \textit{Project Reality} more realistic than the \textit{Battlefield} series and make kills easier. Someone in focus forum pointed out that high TTK solution ``increase weapon stat diversity" but instead characters would not die easily, which is less realistic, because any bullet could be fatal in the real world. Moreover, there is a confusing point worth noting: modifications related to ``instruments" within ``resource files", such as unlimited ammunition and increased damage, can also shorten the TTK. However, altering some specific game characters does not affect the mechanics of the original game, even though these changes often raise issues of competitive fairness. Consequently, the definition of game difficulty presented here is approached from a global or general perspective. Furthermore, too many modifications are potential to change the difficulty, and the detailed explanation will be discussed in Section \ref{subsec_instrument} .

\subsubsection{B.2 The change of game theme}\label{subsec_game_theme}

A new theme often introduces a new storyline or an entirely different setting. \textit{Super Mario Eclipse}\footnote{https://gamebanana.com/MODs/432762} is a prime example, which, as a modification for \textit{Super Mario Sunshine}, includes a user-generated game world within many new levels and challenges. Generally, the form of a new theme varies across different types of games. In open-world role-playing games with high degree of freedom, it often takes the shape of additional stories, such as interacting with special characters to trigger separate side stories. The crucial difference is that a new theme typically requires the creation of a new map, which can also be classified as an object of resource file changes. However, in this paper, we emphasize that only maps that directly alter gameplay are considered as a change in game theme. The distinction between game themes and map variations will be discussed further in Section \ref{subsec_game_object}.

\begin{figure}[h!]
		\includegraphics[width=1.0\columnwidth]{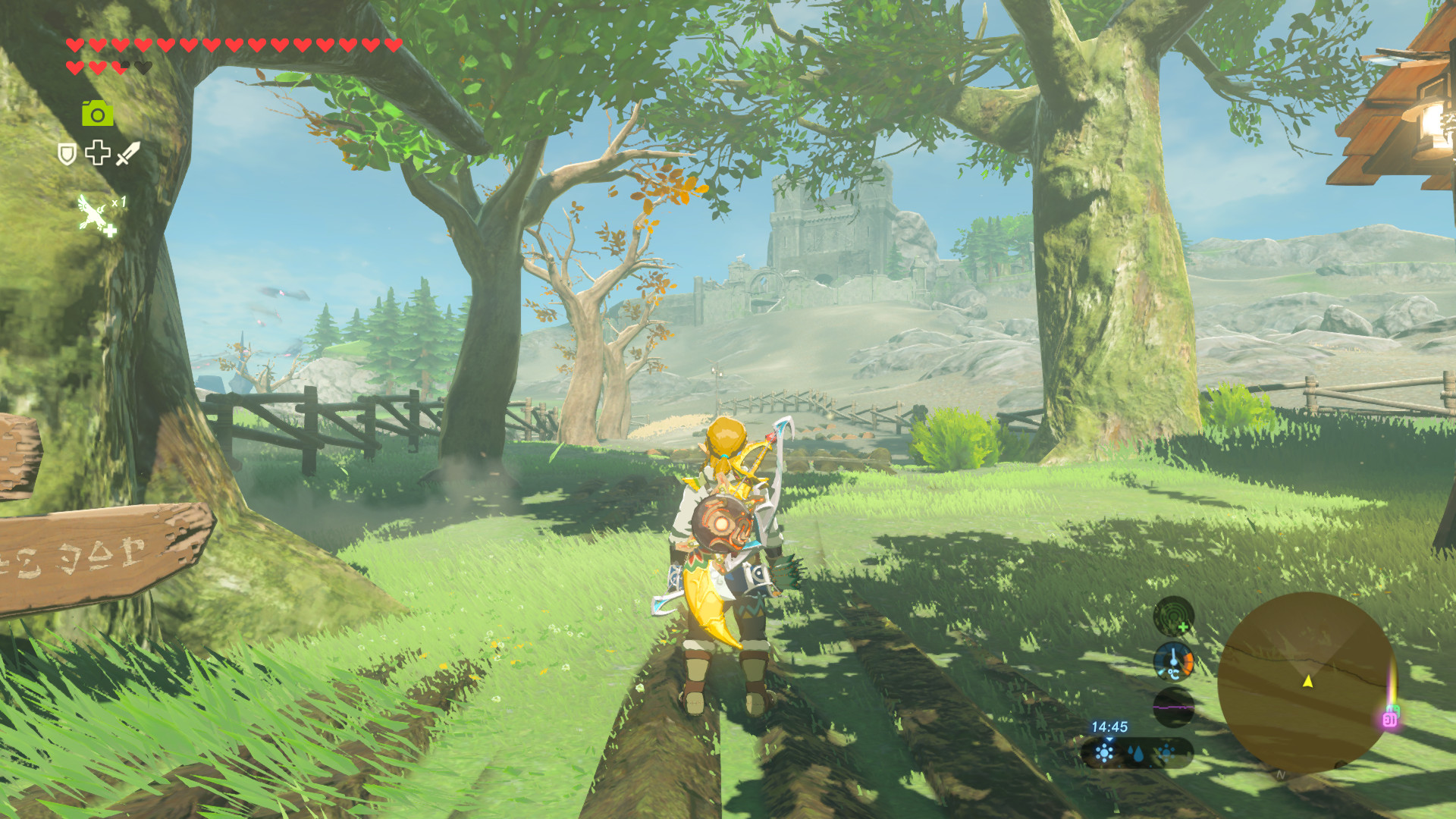}
	\caption{\textit{Second Wind} introduces a new theme for \textit{Breath of the Wild}}
    \label{fig_breath_of_the_wild}
\end{figure}

Another representative example is \textit{Second Wind: The Ancient Trials}\footnote{https://gamebanana.com/projects/35468}, as shown in Figure \ref{fig_breath_of_the_wild}, a project for \textit{The Legend of Zelda: Breath of the Wild}\footnote{https://zelda.nintendo.com/breath-of-the-wild/}. This unofficial expansion pack has garnered significant attention from players eagerly awaiting the sequel of \textit{Breath of the Wild}. To offer something new to players who have completed the original game, the developers introduced new areas, quests, and characters, enriching the game's themes and providing fresh experiences. Different from creating new game modes, such expansion pack is similar to ``Downloadable Content (DLC)", which is additional content distributed by the game's publisher. It is important to distinguish between a new game theme and original game creation, as discussed in Section \ref{subsec_original_game_creation}. The crucial difference lies in the fact that a new theme remains dependent on the framework of the original game, even if it appears to have the characteristics of an independent game. Without the foundational support of the original game, these expansions cannot function on their own. From a purpose-driven perspective, players create expansion packs not to develop standalone games but to extend their experience within a game they’ve likely spent hundreds of hours playing.

\subsubsection{B.3 The change of game path}\label{subsec_game_path}

Here we define the word ``path" as the approach/route to achieve the game target. A game may offer multiple paths to win, e.g., in the bomb defusal mode of \textit{Counter-Strike}, terrorist can win the game either by eliminating all enemies or by planting a C4 bomb in the limited time. In this case, we consider that ``kill all enemies" and ``let the bomb explode" as two different paths. However, this is only an explanation to better understand the definition of game path, while the change of game path we study in this paper does not involve the change of game operation. We propose that a new path should closely resemble the original one. For instance, in an adventure game where the primary objective is to kill the king to achieve victory, an alternative path might involve killing the queen instead. Both paths essentially rely on killing specific enemy to win, and thus, we categorize this as a change in the game's path rather than an alteration to its operation. Changing the game path also has less impact on the original game than introducing new themes, and games that support path modifications often include an editor that allows players to select or create different paths.

\begin{figure}[h!]
	\centering
	\begin{minipage}[t]{0.49\columnwidth}
		\centering
		\includegraphics[width=\columnwidth]{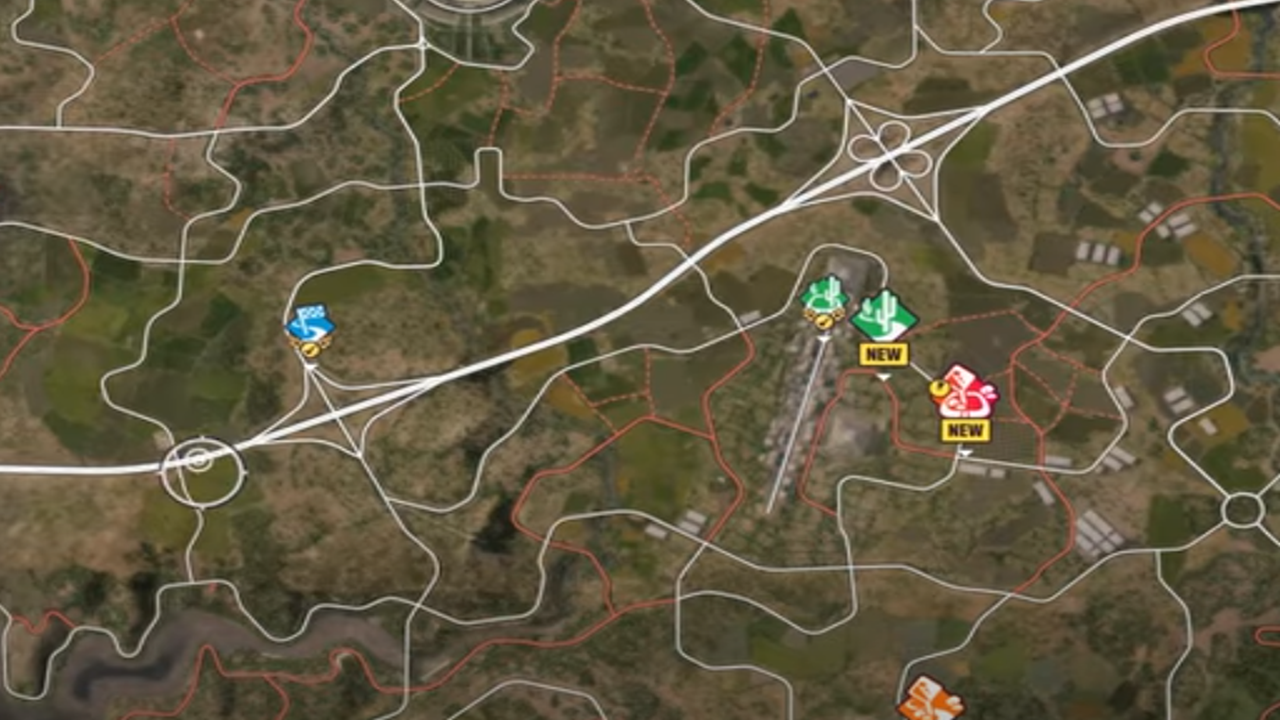}
		\subcaption{EventLab in \textit{Forza Horizon 5}}
	\end{minipage}
	\begin{minipage}[t]{0.49\columnwidth}
		\centering
		\includegraphics[width=\columnwidth]{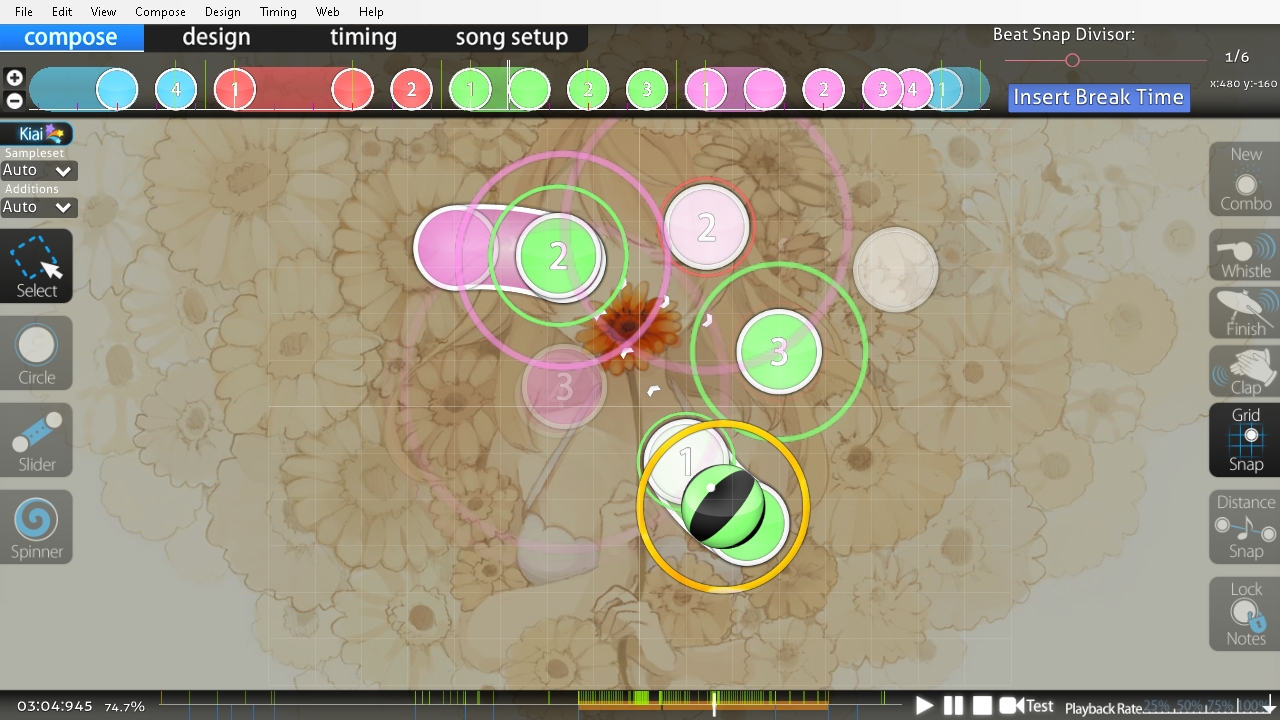}
		\subcaption{Beatmap editor in \textit{Osu!}}
	\end{minipage}
	\caption{How to create new game path}
	\label{fig_game_path}
\end{figure}

The most intuitive embodiment is that in the racing game for the same destination can create different tracks. \textit{Forza Horizon 5}\footnote{https://forza.net/horizon} , a racing game set in Mexico, has the largest map in the entire \textit{Horizon} series \cite{enwiki:1238577496}. Unlike its predecessors, the game introduced \textit{EventLab} to provide a portable toolset for players to create custom tournaments, as shown in Figure \ref{fig_game_path} (a). Since the game is based in an open world environment, there are many different routes from the same starting point to the destination. By marking checkpoints and adding obstacles, players can design and generate new routes.

Another example is rhythm game, in which there are different beatmaps for the same song depending on the difficulty. But users can also create their own beatmaps using the external songs or designing more intriguing paths campared with the original one. As shown in Figure \ref{fig_game_path} (b), \textit{Osu!}\footnote{https://osu.ppy.sh/} supports users to import music from outside and make a beatmap by the built-in editor. In terms of the game rules, changing beatmaps does not change the basic mechanics, while players essentially still need to score points by hitting notes in circles with the mouse. However, since users can create new beatmaps based on their personal understandings of the rhythm and drumbeat of the background music, they can also be regarded as a change of the game path.

\subsection{User-generated instruments}\label{subsec_instrument}

User-generated instrument is a common form of modification, as we discussed in Section \ref{subsubsec_game_difficulty}. These modifications can be used to either skip repetitive or monotonous levels, or to adjust the game's complexity. Unlike changes aimed at altering overall game difficulty, user-generated instruments typically target specific parts of the game rather than the entire gameplay experience. For example, in \textit{Hogwarts Legacy}\footnote{https://www.hogwartslegacy.com}, using a spell to unlock certain rooms still requires solving simple puzzles, a process that can become tedious with repetition. In this case, some players have made modifications to skip the unlocking mechanics, which greatly saves time and optimizes the player's game experience. 

\begin{figure}[h!]
	\centering
		\includegraphics[width=\columnwidth]{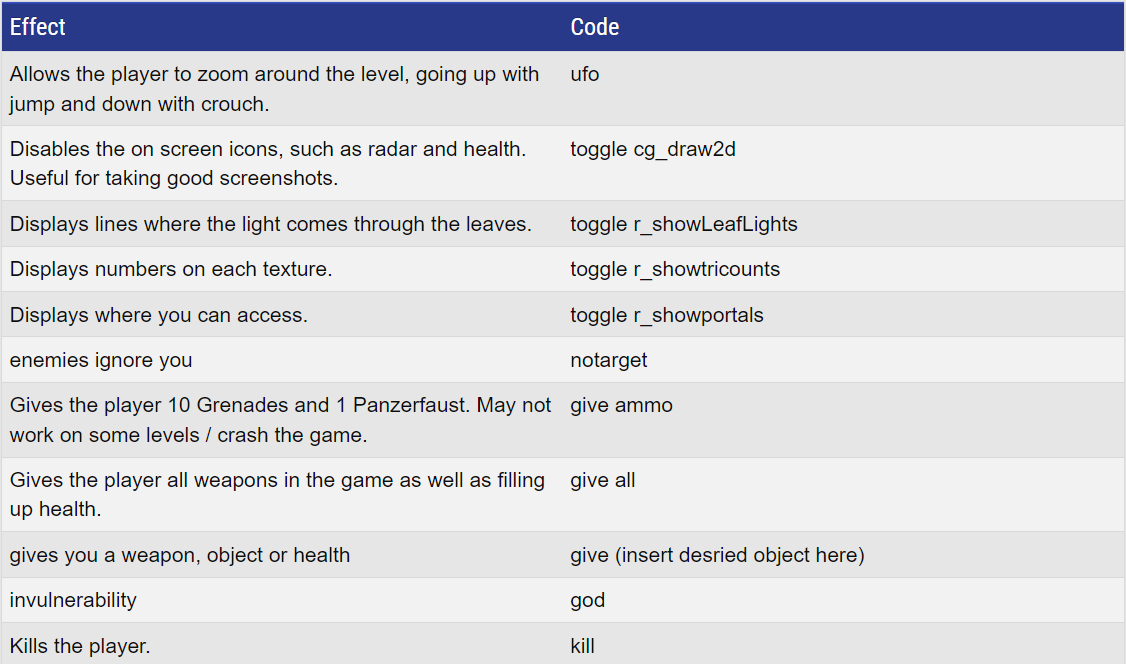}
	\caption{Well-Known Cheating Codes in \textit{Call of Duty} }
	\label{fig_instruments}
\end{figure}

However, many user-generated instruments have the potential to disrupt the game balance in order to improve individual game experience. These tools, often referred to as ``plugins", are frequently used as cheats, particularly in multiplayer environments. Popular modifications like ``Auto Target Enemies” reduce TTK, creating an unfair advantage over other players. In some competitive games, such as \textit{Call of Duty}'s\footnote{https://www.callofduty.com} multiplayer mode, the use of cheat ``plugins" is very common. Figure \ref{fig_instruments} shows some well-known cheating codes which modify the properties of program files to attain desired effects\footnote{https://gamefaqs.gamespot.com/pc/914586-call-of-duty/cheats}. Game developers often crack down on the use of such cheating tools, but we consider that these ``plugins" are indeed user-generated instruments. The abuse of these ``plugins" involves ethical problems, where the relationship between the existence of plugin cheats and game fairness will be further discussed in detail in Section \ref{subsec_ethical}.

In some cases, excessive use of user-generated instruments--i.e., an overabundance of modifications--can lead to qualitative changes that alter the game’s core mechanics. For instance, in a standard competitive shooting mode where players are allowed to respawn at certain points after death, adding a ``no resurrection" modification along with an increase in weapon damage could transform the normal mode into an only-one-life mode, resulting in a more intense game tempo. It is important to note that, in this paper, separating the instruments and game modes does not mean that the two are completely opposed. 

\subsection{User-generated objects}\label{subsec_game_object}

The definition of objects is very broad, where we consider that game characters, items, maps, etc., as resource files of the game, can be regarded as game objects. In most game MODs, altering or adding game objects is one of the most common forms of modification. Therein, one of the most typical changes of game objects is the personalization of characters. Many role-playing games allow players to customize their character’s appearance, including aspects like body shape, skin tone, and facial features, offering a tailored gaming experience that reflects individual preferences.

\begin{figure}[h!]
	\centering
	\begin{minipage}[t]{0.49\columnwidth}
		\centering
		\includegraphics[width=\columnwidth]{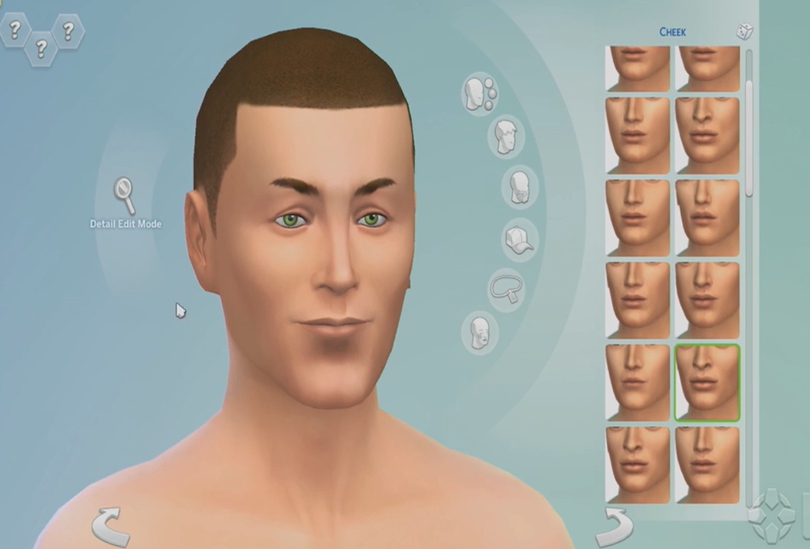}
		\subcaption{Built-in editor in \textit{Sims 4}}
	\end{minipage}
	\begin{minipage}[t]{0.49\columnwidth}
		\centering
		\includegraphics[width=\columnwidth]{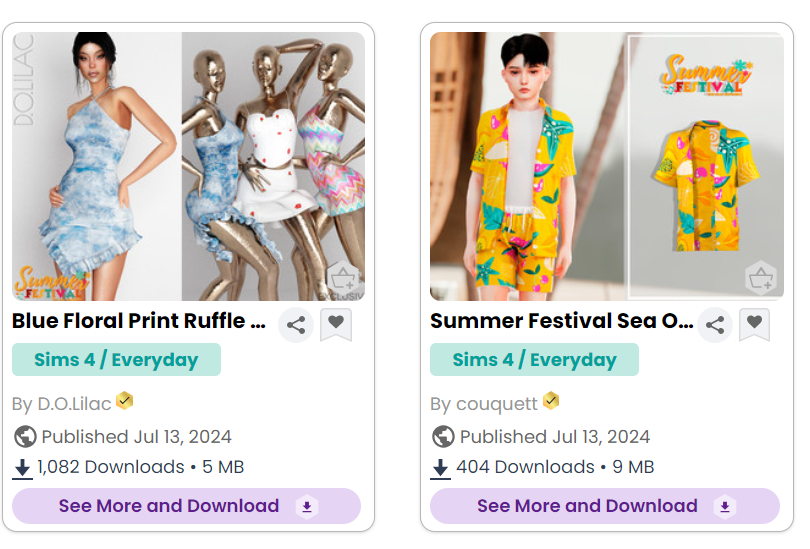}
		\subcaption{External game objects available in game community}
	\end{minipage}
    	\caption{User-generated game objects in \textit{Sims 4}}
	\label{fig_game_object}
\end{figure}

The game \textit{Sims 4}\footnote{https://www.ea.com/games/the-sims/the-sims-4} leverages this characteristic as a major selling point, giving players the freedom to create unique characters and manipulate them to follow the personalized storylines. Even though the developers consistently release expansion packs with new themes such as seasons and high school years, many players are still not satisfied with the existing resources provided by the official developers. MODs address this gap by providing a vast array of additional content created by the community. Notably, platforms like \textit{Nexus MODs} offer a variety of complete modifications for players to download. As shown in Figure \ref{fig_game_object} (b), new clothing will be added by the corresponding MOD, besides there also exist many modifications of houses, furniture and other objects. Like the unique characters generated by the player through the built-in editor (Figure \ref{fig_game_object} (a)), these external resources introduced by MODs are also parts of user-generated objects.

In addition to common objects, there is a special case---maps. As we mentioned above in Section \ref{subsec_game_theme}, maps are an indispensable part of most video games and play a great role. Here we use \textit{Minecraft} as a prototype to make a distinction: If a map serves only as a visual or architectural showcase without embedded game rules--such as a structure designed purely for viewing--it should be categorized as a resource file. However, if a map incorporates interactive elements, such as a secret escape room or a detailed storyline and background, it should be considered as a gameplay modification. For example, the famous modification \textit{``Twilight Forest"}\footnote{https://github.com/TeamTwilight/twilightforest}, which introduces a forest world shrouded in eternal night allows players to beat several powerful bosses as adventurers to earn rewards. While game objects are often integral to gameplay modifications, they are fundamentally types of resource files. Their significance primarily depends on their impact on gameplay, as not all user-generated object changes directly alter gameplay dynamics.

\subsection{User-generated global impacts}

To begin, we must clarify the definition of user-generated global impacts, as it can often be confused with changes to game themes or user-generated maps. A global impact can be defined as a modification that alters the player’s sensory experience--specifically visual, auditory, or tactile sensations--without affecting gameplay mechanics or the original resource files. In essence, global impacts are modifications that adjust how the game feels rather than how it functions. Generally, we conclude three main sensory involvements of players, including visual impacts, auditory impacts, and physical manipulation. In the following sections, we will examine these global impacts individually, using representative examples to illustrate each type.

\subsubsection{E.1 Visual impacts}

\begin{figure}[h!]
	\centering
	\begin{minipage}[t]{0.49\columnwidth}
		\centering
		\includegraphics[width=\columnwidth]{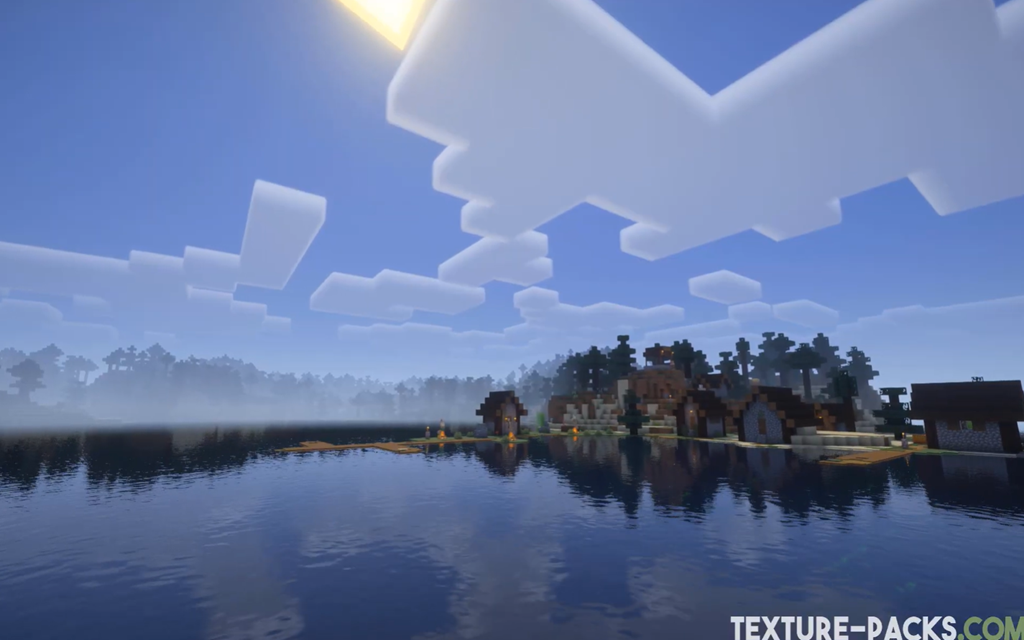}
	\end{minipage}
	\begin{minipage}[t]{0.49\columnwidth}
		\centering
		\includegraphics[width=\columnwidth]{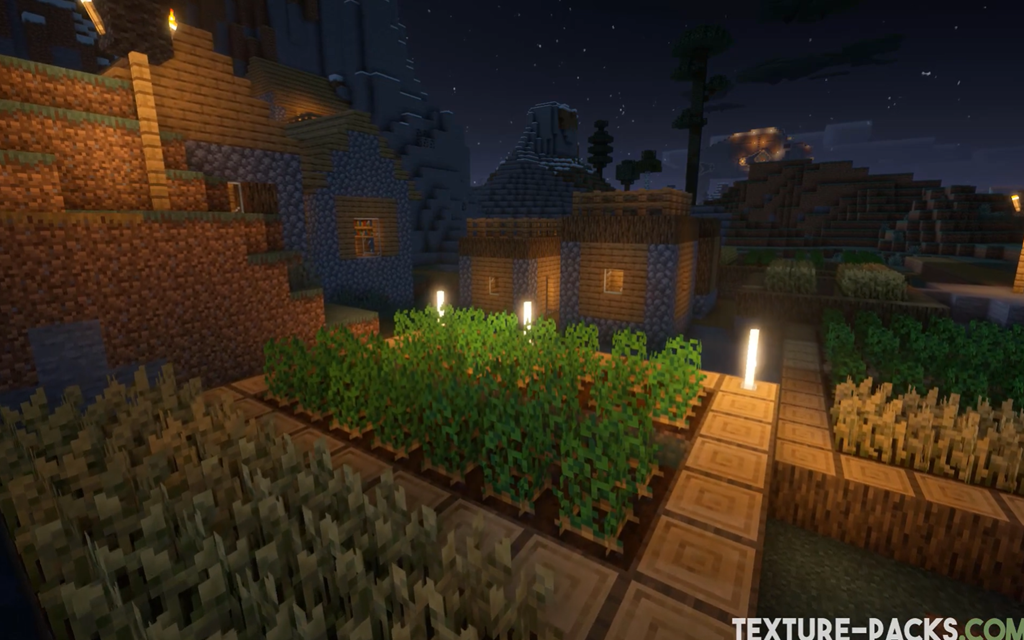}
	\end{minipage}
        \caption{Visual Effect of \textit{Vanilla Plus Shader}\protect\footnotemark}
	\label{fig_visual_impact}
\end{figure}

Visual impact changes are among the most intuitive user-generated modifications, typically altering the appearance of resource files. These changes range from simple color adjustments to complex texture overhauls. The most typical example is \textit{Minecraft}, user-generated texture packs and shader packs contain textures and light change, improving the visual quality of game objects. Figure \ref{fig_visual_impact} shows a famous shader pack named \textit{``Vanilla Plus Shader"}\footnote{https://rre36.com/vanilla-plus} which was developed by ``RRe36". This shader pack enhances lighting, shadow, water effects, plant dynamics, and block materials, all while preserving the original aesthetic of vanilla \textit{Minecraft}. In games that prioritize immersive visuals less, filter modifications are more prevalent. For instance, MODs that remove the default dark green filter in \textit{Starfield}\footnote{https://bethesda.net/en/game/starfield} clarify the visual display, improving overall screen visibility. Additionally, certain filters can enhance the gaming experience for color-blind players, allowing for greater visual accessibility.

In 2022, Lin \etal \cite{lin2022features} proposed using ``impact feel" to describe how players feel when they receive feedback from juicy impact in a game. We consider that such feedback is a combination of visual and auditory effects. In their study, they trained a Natural Language Processing (NLP) model to rank the impact feel of action games, finding that hit stop, sound coherence, and camera control may significantly influence the impact feel. For action game players, enhancing attack effects--such as increasing color contrast, adjusting view angles, and intensifying visual details--falls within the scope of global visual impacts. Although few current UGC projects focus on enhancing impact feel, its importance in player experience suggests it may become a valuable area in future game development.

\subsubsection{E.2 Auditory impacts}

We consider that background music, character voices, and all sound effects should be auditory impacts. Some games allow players to upload music files to play (while there might exist copyright issues in practice), or even replace the original sound effects with another game's sounds. A typical examples can be found in \textit{Palworld} \footnote{https://www.pocketpair.jp/palworld}. The MOD \textit{``Ninja Slayer Audio"}\footnote{https://www.nexusmods.com/palworld/MODs/771} created by sg12041 substitutes the voices of player and human enemy with dialogue from the \textit{Ninja Slayer} animation. Sometimes a game designer's intention for a particular sound is at odds with the player's experience. Many players complained that the sounds made by Lucky Pals were annoying when playing the game, so a modification called \textit{``Shiny Lucky Pals No Sound Shining"}\footnote{https://www.nexusmods.com/palworld/MODs/124} was created to eliminate the sounds made by these creatures. The changes in the game experience brought about by changing the audio are also intuitive, and sometimes even affect the player's manipulation, such as using clearer ballistic sounds to allow the player to judge the position faster.

\subsubsection{E.3 Physical manipulations}

Physical manipulation refers to the methods players use to interact with games, typically through keyboard and mouse input. However, some games--especially those from developers like \textit{Nintendo}\footnote{https://www.nintendo.com} and \textit{PlayStation}\footnote{https://www.playstation.com}--may be limited to specific hardware or lack support for alternative controllers. To enable diverse gameplay experiences, modifications can provide users with different control options. For games such as \textit{Fallout: New Vegas}\footnote{https://fallout.bethesda.net/en/games/fallout-new-vegas} that do not support controller input, MODs can be used to enable gamepad functionality, offering a more comfortable experience for some players. Additionally, specialized input methods, such as using a simulated steering wheel in \textit{Grand Theft Auto V}\footnote{https://www.rockstargames.com/gta-v}, allow for a driving experience closer to real life, enhancing immersion beyond traditional keyboard and mouse controls. These modifications significantly enrich the gameplay experience by providing alternative ways to physically engage with the game world.

\begin{figure}[h!]
	\centering
		\includegraphics[width=\columnwidth]{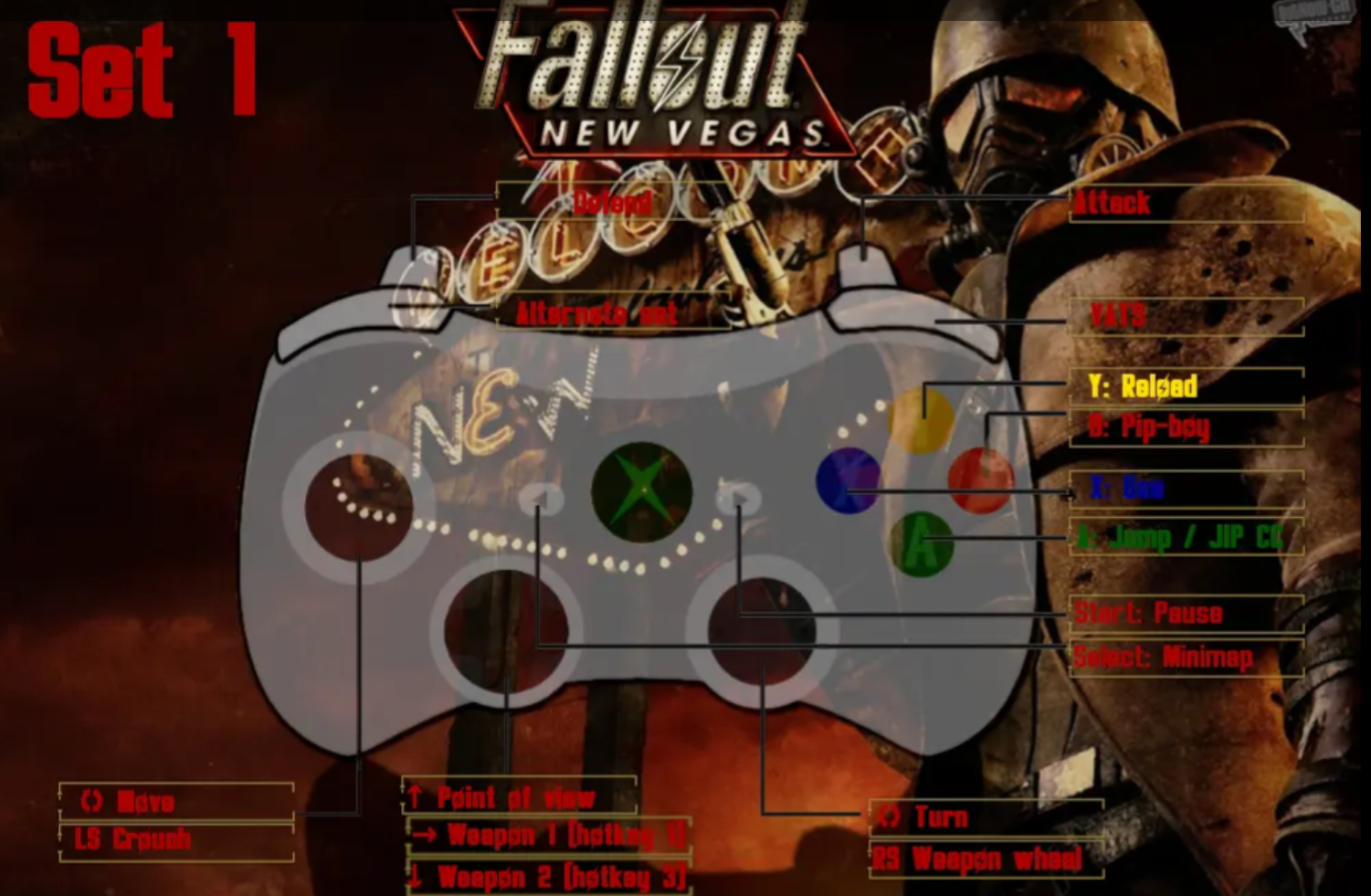}
	\caption{FNV Gamepad For MODs\protect\footnotemark}
	\label{fig_physical_manipulation}
\end{figure}
\footnotetext{https://www.nexusmods.com/newvegas/MODs/70603?tab=images}

\begin{figure*}
    \centering
    \includegraphics[width=2.0\columnwidth]{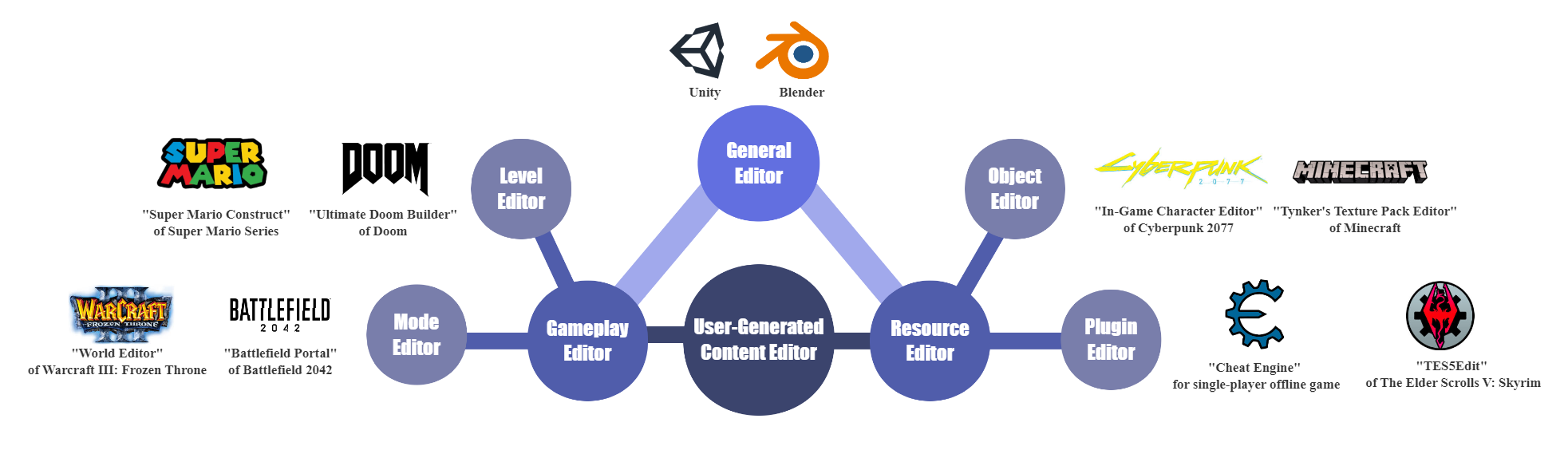}
    \caption{Parallel classification diagram of user-generated content editors}
    \label{fig_editor}
\end{figure*}

\section{User-Generated Content Editor} \label{sec_editor}

A UGC editor is user-centric and serves as the primary tool for creating UGC, typically without requiring users to have advanced programming skills compared with professional game editors \cite{ng2018situated}. There are two different ideas of generation that need to be declared here: (1) The first is professional generation for professional game developers who seek to create fully-fledged indie games; (2) The second is the generation for ordinary users, since most users lack the skills to develop independent games, and the end product of this kind of generation is what we call UGC. Unlike game engines, UGC editors are designed as extensions of games, aimed at providing novel gaming experiences. They enhance replayability and foster community engagement by allowing players to create and share their own content. 

In our conference version \cite{duan2022user}, we divided UGC editors into four categories: level editors, gameplay editors, game editing engines, and community-made game editors, and we grouped them together with UGC as part of the classification Venn diagram. Upon further analysis, we identified a lack of consistent criteria in this classification. For example, community-made game editors were by definition only intended to distinguish them from official game editors, but they were not well differentiated from other genres in the UGC they created. In other words, a community-made editor might be involved in gameplay or level generation. The original article used the same decision-tree classification as UGC to deal with this problem, with ``official vs. unofficial" as the highest judgment priority. However, such classification inevitably caused the community-made game editor to be unable to continue to distinguish further down, according to the UGC created where we believe should be mainly focused on. Besides, the original taxonomy neglected game resource editors, which were often the most common way to create UGC, such as the character shape editors provided by some role-playing games that did not belong to any category in the original classification. In conclusion, we believe that the original classification in our conference version \cite{duan2022user} is inadequate for accurately summarizing the current state of game UGC editors, as it lacks the rigor necessity to effectively categorize these diverse tools.

Therefore, in this paper, we ignore whether there is any participation of the official game developers and use UGC as a unified classification standard to build another similar parallel classification bubble diagram for UGC editors as shown in the Figure \ref{fig_editor}, which contains two types, including gameplay editors and resource editors. The details about each category of UGC editors will be discussed in the following subsections. 

Additionally, we introduce the concept of ``general editor" to provide a dialectical comparison with UGC editors. While common game development tools like \textit{Unity}\footnote{https://unity.com} and \textit{Blender}\footnote{https://www.blender.org/} have been increasingly used to create UGC in recent years, they do not fit neatly into any specific category based on the classification criteria for UGC editors outlined in this article. Despite this, the content produced by users through these general editors can be categorized as UGC, and thus, the corresponding tools may be considered UGC editors under this broader classification.

\subsection{Gameplay Editors}

As we mentioned in Section \ref{sec_ugc}, for UGC, changing the default rules or adding new levels is a change in gameplay. Based on this definition, we name the tools that can provide such UGC as the gameplay editors. Then, we continue to subdivide them into mode editors and level editors according to the different UGC created by there editors.

\subsubsection{Mode Editors} \label{subsec_mode_editors}

\begin{figure}[h!]
	\centering
		\includegraphics[width=\columnwidth]{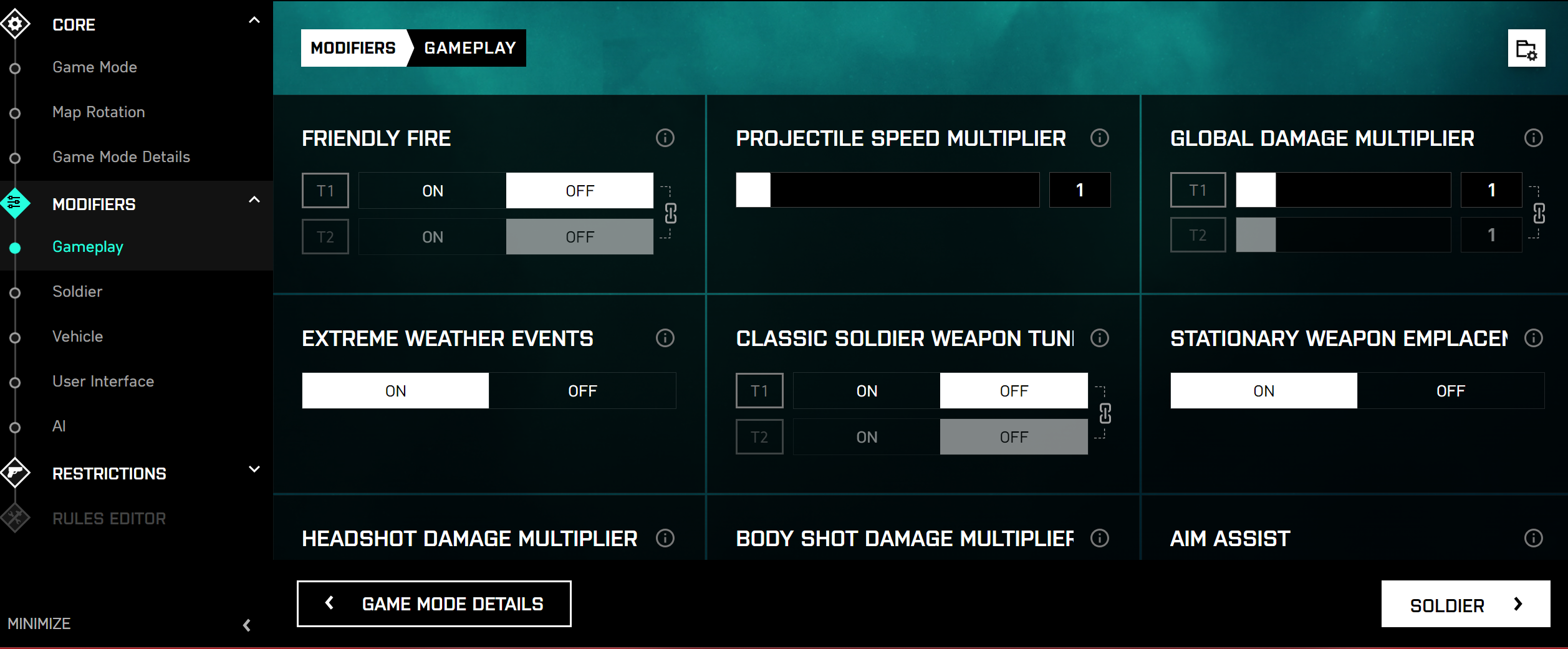}
	\caption{The official mode editor gives players a high degree of freedom}
	\label{fig_mode_editor}
\end{figure}

Some AAA (3A) games offer customizable mode editors that allow players to modify the game mode and basic rules, such as the number of people in each match and the limited time. A typical example is \textit{Battlefield Portal}\footnote{https://www.ea.com/games/battlefield/battlefield-2042/game-overview/modes/battlefield-portal} released by \textit{Battlefield 2042}\footnote{https://www.ea.com/games/battlefield/battlefield-2042}, which allows players to create new game modes. Figure \ref{fig_mode_editor} shows the editor interface on the official website of \textit{Battlefield 2042}. This editor clearly empowers the player to establish the game’s rules. From selecting the mode and choosing the map to defining specific gameplay rules and adjusting character and weapon modifiers, the player enjoys significant freedom in customizing the experience. This is the core function of what we define as a mode editor.

Another representative example is \textit{Warcraft III: Frozen Throne} which includes the \textit{World Editor}, an official level editor released by \textit{Blizzard Entertainment}\footnote{https://www.blizzard.com/en-us/}. The \textit{World Editor} allows users to design custom maps, create new units, set up triggers, and even script custom game mechanics. This powerful tool enables players, even without programming expertise, to easily create custom game modes. Some of the most popular custom games, like \textit{DOTA} and \textit{The Lord of the Rings}, originated from the \textit{Warcraft III World Editor}. Although we classified the \textit{World Editor} as a level editor in our conference version \cite{duan2022user}, it was intended to encourage players to design new maps and generate new campaigns. Therefore, we believe that editors giving players the power to reset the rules of the game should be identified as a mode editor, since it is our fundamental criterion to distinguish mode editors from level editors.

In general, creating new game modes is often more challenging than designing new levels within existing rules, as developers must carefully consider the balance of the scripts running in the game and ensure the editor's functions are both robust and precise. To minimize the likelihood of bugs, the significant potential of game mode editors comes with equally high development costs, largely due to the complexity involved in creating these advanced tools. This also explains why gameplay editors are more commonly found in 3A titles rather than indie games. Furthermore, modifying a game mode often involves altering the core mechanisms of the game, which typically requires writing game scripts and modifying the code. Unless there exist an official editor provided by the developers, it is challenging for players without a programming background to create new game modes. This is one reason why editors that allow for game modes modification are relatively rare. Additionally, creating and releasing such editors as unofficial tools can raise issues related to copyright and conflicts of interest. For example, as an open source third-party game editor, \textit{HiveWE}\footnote{https://www.hiveworkshop.com/threads/hivewe-world-editor-0-6.303110/} is not officially supported by \textit{Warcraft III} and cannot distribute the editor for legal reasons. We will further elaborate in Section \ref{subsec_guc} on whether authorities should provide adequate support to third parties and pursue cooperative interests.

\subsubsection{Level Editors}\label{subsec_level editor}

Level generation, a key challenge in Procedural Content Generation (PCG), has emerged as a significant research topic in recent years \cite{khalifa2016general}. The concept of level editor has been well-established in recent years, with many adventure and RTS games offering tools for level design. Ipacs \cite{Ipacs_2024} suggests that levels are not merely a series of obstacles but rather thoughtful combinations of elements designed to provide players with new experiences, but she ignores the contradiction of whether the level should obey the rules of the game. In the context of UGC, it is significant to differentiate between levels and modes. Some editors, marketed as ``level editors" (such as the \textit{World Editor} mentioned in Section \ref{subsec_mode_editors}), may not strictly fit this classification, as they offer users extensive freedom, enabling them to produce content that may deviate from the core mechanics of the original game.

We define a level editor as a game editor that enables users to use existing elements to create new levels or campaigns using existing in-game elements without changing the game's rules or targets. Unlike mode editors, which modify the core structure of the game, level editors maintain the fundamental framework, making them more affordable to develop and widely accessible. For example, as we mentioned in Section \ref{subsec_game_path}, using \textit{EventLab} to add obstacles and design a new track in \textit{Forza Horizon 5} could be considered a new user-generated level. \textit{Doom}\footnote{https://bethesda.net/en/game/doom} supports various level editors, and the latest \textit{Ultimate Doom Builder}\footnote{https://ultimatedoombuilder.github.io/} includes multiple new features for players to edit levels in addition to the previous \textit{GZDoom Builder}\footnote{https://github.com/m-x-d/GZDoom-Builder} features.

Creating new levels can revitalize an existing game and significantly extend its lifespan by offering users creative freedom. Communities built around these editors provide a platform for sharing and collaboration. For example, \textit{Level Share Square}\footnote{https://levelsharesquare.com/} is a level design community unaffiliated with any specific Intellectual Property (IP) that currently supports games from the \textit{Super Mario series}. Figure \ref{subsec_level editor} shows an online level editor called \textit{Super Mario Construct}\footnote{https://levelsharesquare.com/games/supermarioconstruct}, with a concise editing page. In general, a level editor typically provides users with predefined in-game elements, limiting the extent of customization. For instance, users can modify aspects like the color scheme of obstacles, but they generally cannot change models by importing external resources. On the other hand, the combination of browser-based editor and community is not uncommon, some 2D adventure games seem to be more suitable for such type of level editor, which does not require to download additional software, reducing the development cost.

Recently, with advancements in AI, the integration of AI in level design has increasingly become a focal point of interest. Guzdial \etal \cite{guzdial2019friend} designed an AI-driven game level editor named ``\textit{Morai Maker}" for \textit{Super Mario Bros.} to design style games and gave a deep insight into the relationship between AI and level designers. The integration of AI into game level editors may revolutionize how users interact with and create content in games, since AI can act as a co-creator, offering suggestions, optimizing design elements, and even generating entire levels based on minimal input from users, or analyze a player's gameplay style and automatically generate levels that challenge and engage them based on their unique play patterns \cite{guzdial2016game, summerville2018procedural}. For both novice and experienced users, AI can make level customization much easier, even if there are issues with the fairness of creation and the potential uncontrollability of AI-Generated Content (AIGC)\cite{guzdial2018automated,cook2014ludus}.

\begin{figure}[h!]
	\centering
	\begin{minipage}[t]{0.49\columnwidth}
		\centering
		\includegraphics[width=\columnwidth]{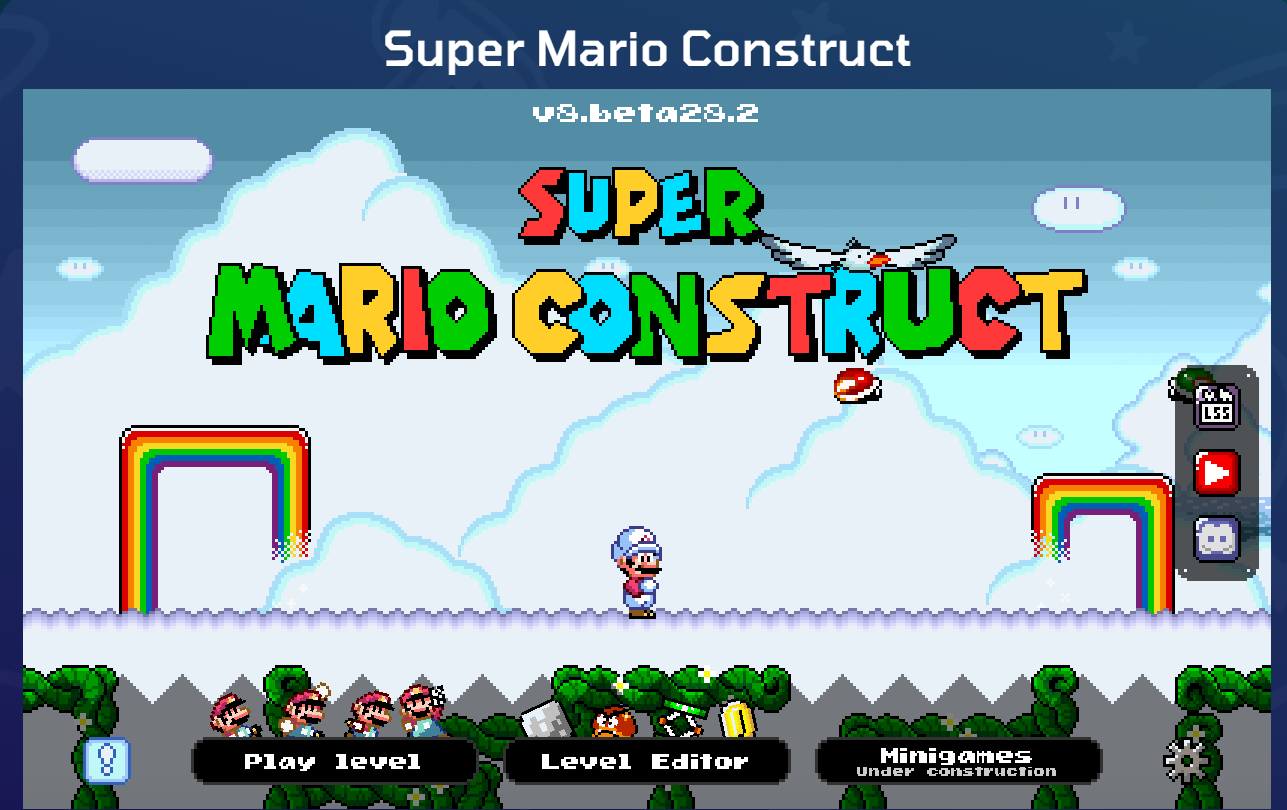}
	\end{minipage}
	\begin{minipage}[t]{0.49\columnwidth}
		\centering
		\includegraphics[width=\columnwidth]{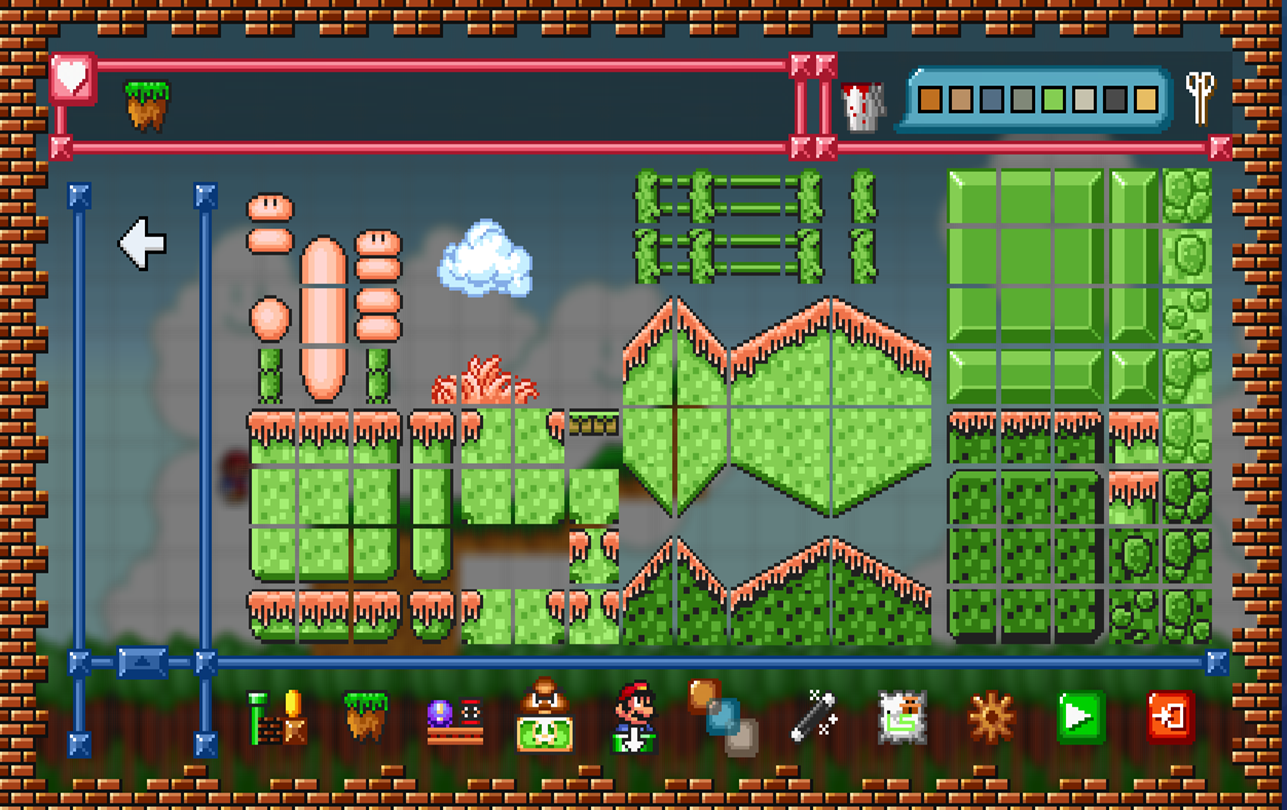}
	\end{minipage}
        \caption{Level editor in \textit{Super Mario Construct}}
	\label{fig_level_editor}
\end{figure}

\subsection{Resource Editors}

Unlike the original classification, we introduce the concept of ``resource editor", because the resource files that can be changed or generated by users in video games also fall under the category of UGC. Generally speaking, we classify the resource editors into two directions, including object editors and plugin editors.

\subsubsection{Object Editors}

Object editors are the most common game editors. Out conference version \cite{duan2022user} overlooked the presence of functions that allow customization of game objects in certain video games, which we now classify as object editors, even though they are integrated into the game itself. Role-playing games often feature fully functional and detailed character editors, but sometimes the default appearance of the character is not satisfactory to all players. Moreover, for the main character that the player primarily controls, if the developer can provide the corresponding editor, it can solve this contradiction to a certain extent. \textit{Cyberpunk 2077}\footnote{https://www.cyberpunk.net/hk/en/} features a wide range of customizable character traits, from gender to body type to facial features, that can be modified through the in-game character editor. These official object editors sometimes even indirectly affect the entire game play experience, such as \textit{Sims 4} mentioned in \ref{subsec_game_object}. To this end, the functions of the in-game editor may also become one of the factors to evaluate the quality of the game. 

In addition to official in-game editors, there are fully functional unofficial modification tools. For instance, \textit{Tynker's texture pack editor}\footnote{https://www.tynker.com/minecraft/items/edit/} is an unofficial object editor for \textit{Minecraft} that allows users to edit pixels on the website to change different game items, character skins, and blocks. In contrast to the official editor, the resource files generated in the unofficial editor usually need to be mannually imported into the game by the user. Moreover, compared with other types of editors, object editor essentially adds or replaces the original resource file (such as creating a new character model file) to achieve the purpose of modification, so for object editors, the difference between official and unofficial is not that obvious. Unofficial developers typically face challenges in accessing core game files and are often restricted by official developers from altering gameplay mechanics. The use of object editors is generally similar to model building, and in terms of functionality, general editors such as \textit{Blender} can also achieve sophisticated modeling. Using external modeling software to create more detailed models for importing into a game is common practice. For instance, many user-made MODs in \textit{Sims 4} are created this way. In \textit{Roblox}, these externally imported 3D models are known as ``meshes", although players can also create models directly in \textit{Roblox Studio}, the advantage of using external general editors lies in their greater professionalism and power.

\begin{figure}[h!]
	\centering
	\begin{minipage}[t]{0.49\columnwidth}
		\centering
		\includegraphics[width=\columnwidth]{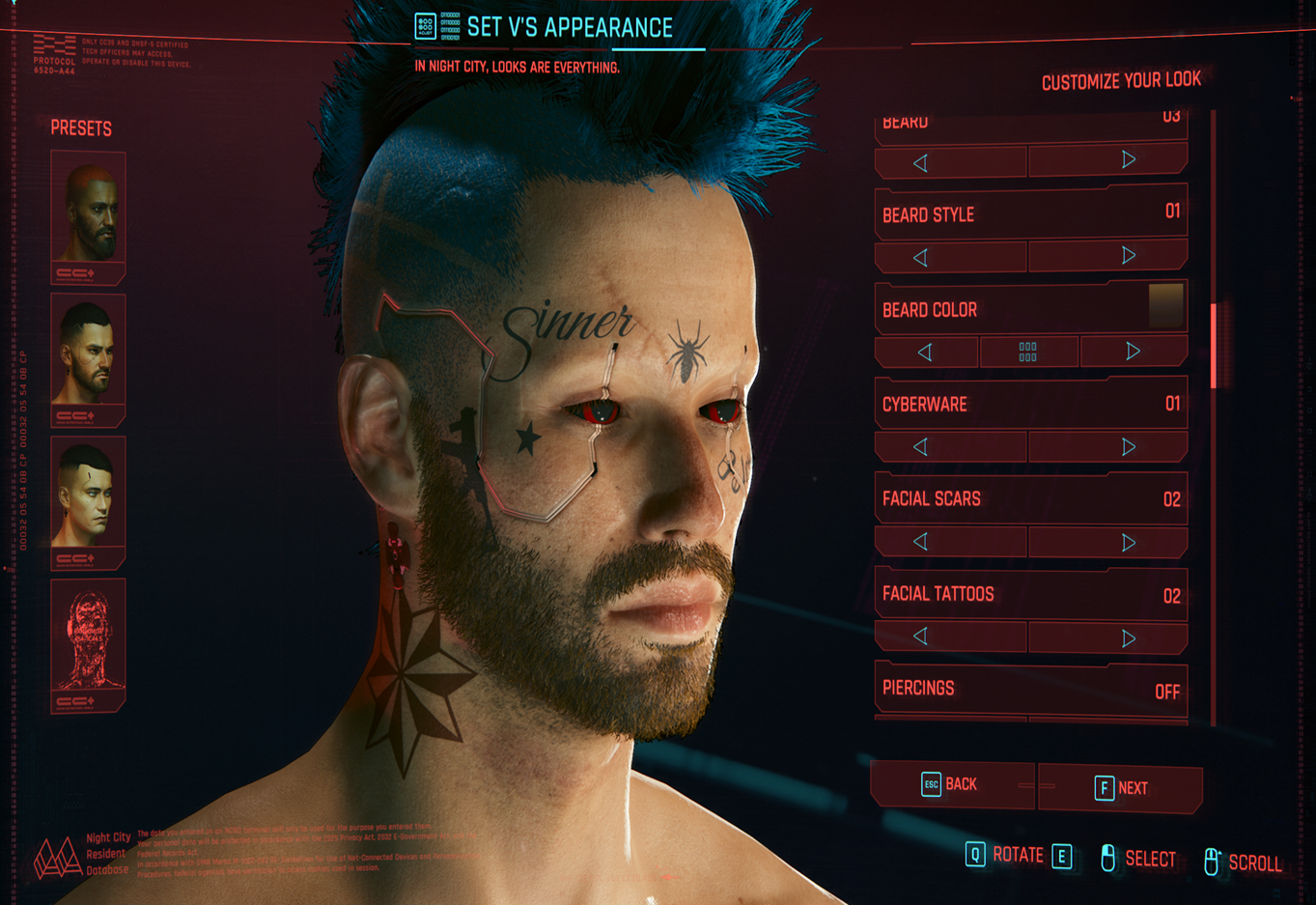}
	\end{minipage}
	\begin{minipage}[t]{0.49\columnwidth}
		\centering
		\includegraphics[width=\columnwidth]{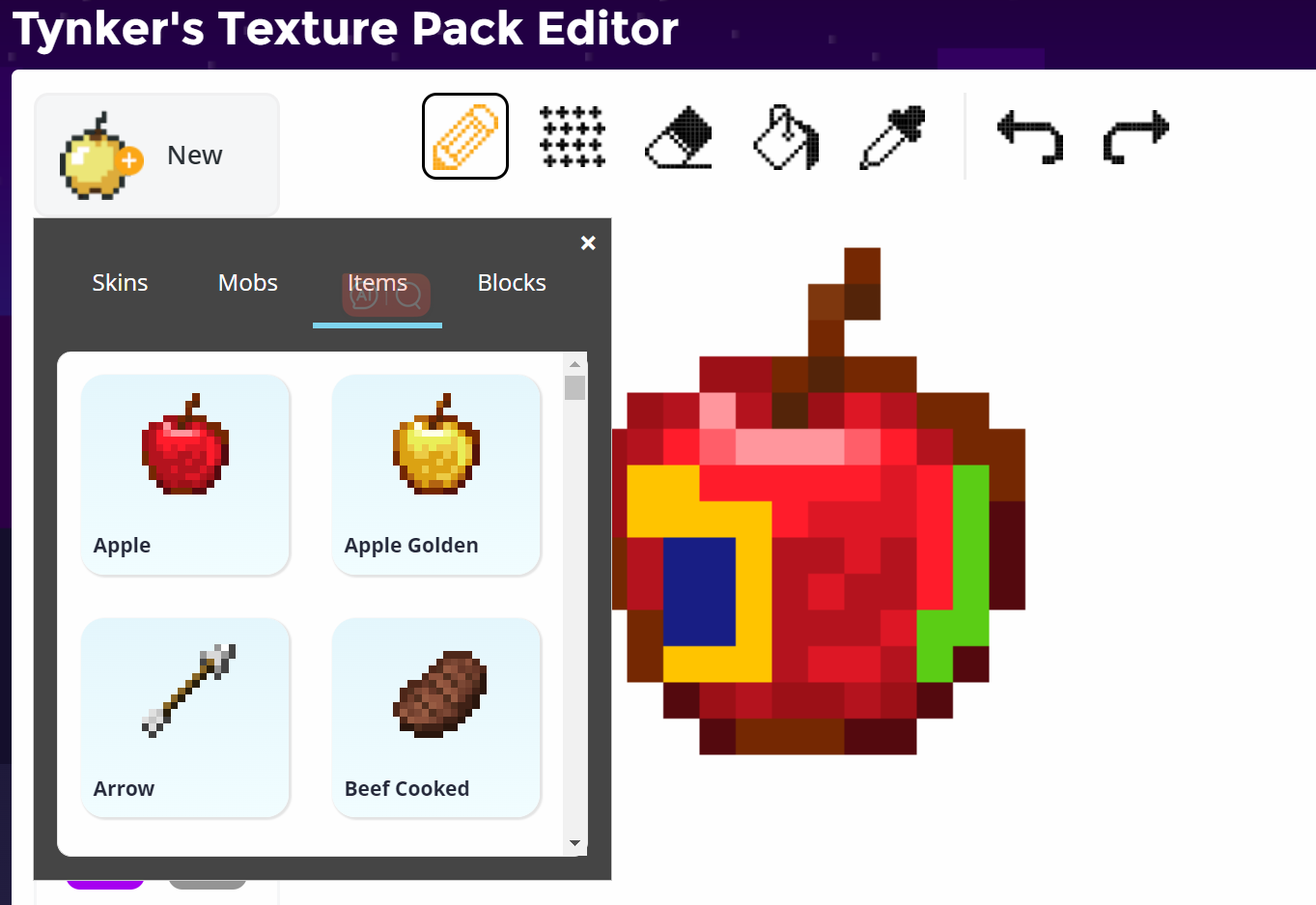}
	\end{minipage}
        \caption{Comparison of built-in and unofficial object editors}
	\label{fig_object_editor}
\end{figure}

\subsubsection{Plugin Editors}

Plugin editor is aimed to generate or manage plugins that can be applied in the game. The definition of the plugins has been described in Section \ref{subsec_instrument}, which corresponds to the user-generated instrument. Even if we classify plugins as a type of resource file, the plugin editor is much more complex than object editor, and usually involves accessing the core file and modifying some data of its settings. So far, the plugin editor only accounts for a small proportion of UGC editors, perhaps because creating plugins that can be applied to games requires users to have a certain programming background, and modifying game values also needs to be done through common Integrated Development Environment (IDE) such as \textit{Visual Studio}\footnote{https://visualstudio.microsoft.com/}. However, compilers that can be used to write plugin code files cannot be considered as plugin editors. We believe that the UGC editor should be distinguished from the tools used by professional developers. 

\begin{figure}[h!]
		\includegraphics[width=1.0\columnwidth]{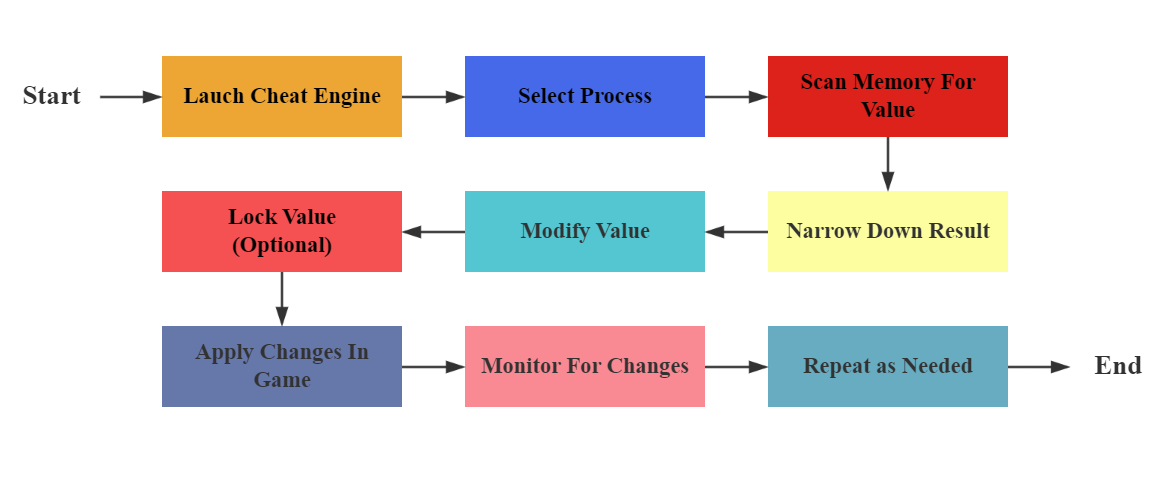}
	\caption{A simplified flowchart of how \textit{Cheat Engine} works when modifying a game's memory}
    \label{fig_cheat_engine}
\end{figure}

\textit{Cheat Engine}\footnote{https://www.cheatengine.org/index.php} is a modification tool for single-player games without internet connection. It allows users to modify some game parameters (such as health and damage) according to their needs. Equipped with a memory scanner, \textit{Cheat Engine} enables users to find the correct variable and make numerical changes to debug them by quickly scanning the code files, where Figure \ref{fig_cheat_engine} illustrates a simplified flowchart of modifying a game's memory. For example, given that the initial Health Points (HP) was 100 and the health was reduced by 1 after a hit. The \textit{Cheat Engine} will scan for a variable with a value of 100, and then use the simulated ``hit me" function in \textit{Cheat Engine} to look again for a variable with a value of 99, which could narrow down the variables that actually point to HP so that other following changes of value can be made.

\textit{TES5Edit}\footnote{https://github.com/TES5Edit/TES5Edit/releases/} is another special case of plugin editor used to operate and manage \textit{The Elder Scrolls V: Skyrim}\footnote{https://elderscrolls.bethesda.net/en/skyrim10} plugins. In addition to generating plugins, its main function is to compare multiple plugins and find potential conflicts to achieve efficient management. There may be incompatibilities between different MODs, but for players who have multiple MODs installed in the same game, checking one by one is quite a trouble and can even reduce the player's interest in the game. Therefore, tools such as ``module manager" can be used to solve this problem, and \textit{TES5Edit} can also be used to remove official DLC that may cause the program to crash.

However, \textit{Cheat Engine} is only available for offline single-player games and may not workable for online multiplayer games. Both of the above examples are editors for single-player games, while in fact, the usage of plugins in multiplayer games, especially competitive multiplayer games, is an issue of fairness in the game, which is why developers in competitive games often openly prohibit cheating in the game. Online video games involve complex data transfers between servers and untrusted clients \cite{laurens2007novel}, making it harder for plugin developers to acquire code and make changes. Besides, multiplayer games typically feature dedicated anti-cheat systems or administrators tasked with detecting and preventing cheating, with significant penalties for those caught violating the rules. Despite the ongoing issues with cheating in popular FPS games like \textit{Valorant}\footnote{https://playvalorant.com/}, cheating plugins and the possible editors used to create them are generally not publicly available due to the strong opposition from game developers and the implementation of robust anti-cheat measures.

\section{Discussion} \label{sec_dis}

In the previous sections, we completed the classifications of UGC and UGC editors and analyzed relevant examples. However, there still exists limitations in our study. The following subsections will address the issues raised above and outline prospects for future research.

\subsection{Overlooked Role of Fandom of Video Game Culture}

The UGC discussed in this paper mainly focuses on content that directly impacts the original game, with our study examining how UGC alters the game itself. However, it is also important to acknowledge the existence of UGC that extends beyond the game itself, resulting in derivative works such as videos, fan fiction, and other creative outputs inspired by the game's characters and storylines. In recent years, these secondary creations based on existing works have been referred to as ``fandom". As a part of popular culture, fandom by definition fits into Wirman's \cite{wirman2009productivity} five dimensions of player productivity as ``productivity beyond play: expressive productivity", while the UGC we studied in this paper belongs the ``productivity for play : instrumental productivity" \cite{ferreira2022fandom}. Moreover, Jones \cite{jones2006shooting} believes that video games provide a great platform for studying fan culture. Similarly, some researchers have pointed out that video game is a representation of participatory culture \cite{postigo2008video, li2010digital, raessens2005computer, dena2008emerging}, where fans' creative contributions can create many different forms in the video games.

Since this part of UGC falls outside the scope of our classification, we acknowledge that the classification system we propose may not be fully comprehensive. Regarding the fandom of video culture, there are some studies have provided detailed discussions, so we list a part of representative works for reference. For example, Dym \etal \cite{dym2018theyre} pointed out that fan-created communities can broaden the scope of video game culture, introducing more gender diversity in storytelling and challenging hegemonic barriers. Pearce \cite{pearce2011communities} argued that fandom has expanded the culture of video games, as fans not only consume content but also produce, modify, and distribute their creations. Booth \cite{booth2015game} considered that fan-created content also influenced the perception and experience of games themselves. In fact, as an indispensable part of UGC, the influence of fan creation on video game culture and communities continues to grow \cite{gray2017fandom}. In future research, exploring the relationship between fan creations and video games, as well as their impact, will be highly valuable for both industry and academia.

\subsection{Relationship Between Game Genre and User Creative Tendencies}

Our study suggests a potential link between the type of game and the tendency of users to create UGC. For instance, in sandbox games like \textit{Minecraft}, UGC frequently takes the form of resource files, whereas in adventure games such as the \textit{Mario series}, users tend to create new levels or campaigns. The tendency toward UGC seems to be related to the convenience that the games provide for the creation of corresponding genres. Additionally, the attitudes of game developers toward UGC appear to influence users' willingness to engage in content creation. Daugherty \etal \cite{daugherty2008exploring} pointed out that incentives related to content creation may significantly impact user motivation. However, existing research on UGC creation has predominantly focused on user-driven factors, leaving a gap in academic inquiry regarding the relationship between game types and user motivation for UGC creation. Therefore, we also consider that game type should be regarded as an external drive influencing users' propensity to create UGC, with different game genres potentially shaping different characteristic of the UGC produced.

\subsection{Application of Artificial Intelligence in User-Generated Content}

As an advanced form of Procedural Content Generation (PCG), AI-Generated Content (AIGC) has recently been applied in various games, particularly in character design and resource generation \cite{shen2023effects,yannakakis2018artificial,schaeffer2002games}. Currently, AIGC mainly uses deep learning methods, which can be more powerful and flexible than traditional level or map generation methods \cite{shen2023effects}. For instance, Mi \etal \cite{mi2023application} used the Black bear monster from ``Journey to the West" to introduce an application path of AIGC in actual game character design. 

We can utilize \textit{Minecraft}’s world generation as an illustrative example: The traditional method involves providing a seed (typically a string of limited length) through a specific algorithm, generates a pseudo-random value (as computers inherently struggle with generating true randomness) \cite{bergensten2011short}. Therefore, this value of the seed determines the terrain and features of the world. In fact, the process of manually entering the value of the seed in the game to enjoy the random generation of the world can also be regarded as a kind of UGC creation. However, the seed is not directly related to the actual generated world. On the one hand, since the block updates and species generation are dictated by the developer's algorithm, the same seed will consistently produce the same world. On the other hand, a limitation of this approach is that slight variations in the algorithm across different game versions can lead to unpredictable differences in the terrain and landscapes generated within the game. For example, a seed named ``rainforest" does not mean that it can generate a world with rainforest landforms. 

\begin{figure}[h!]
	\centering
	\begin{minipage}[t]{0.49\columnwidth}
		\centering
		\includegraphics[width=\columnwidth]{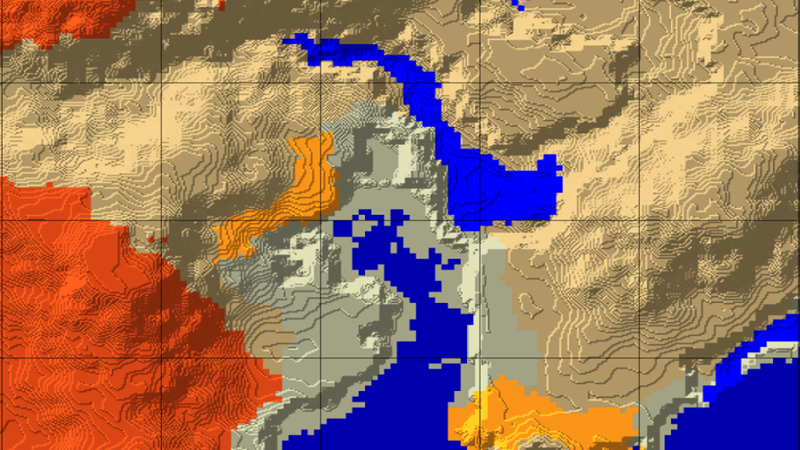}
        \subcaption{Seed-Generated World}
	\end{minipage}
	\begin{minipage}[t]{0.49\columnwidth}
		\centering
		\includegraphics[width=\columnwidth]{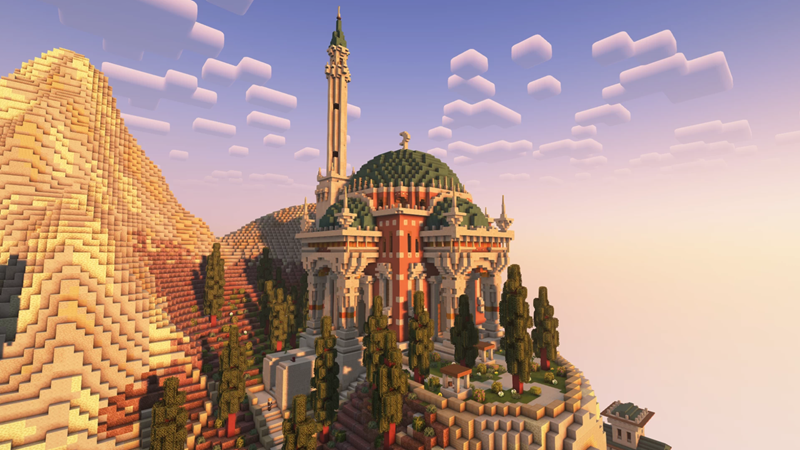}
        \subcaption{AI-Assisted Dune City Created By Kaizen87}
	\end{minipage}
        \caption{Comparison of World Generation in \textit{Minecraft}}
	\label{fig_disscussion_c}
\end{figure}

To meet the requirements, users can choose to customize each block with coordinates in some unofficial map editor such as \textit{MCEdit}\footnote{https://www.mcedit.net/}, but setting blocks one by one can be inefficient, because in the classic version of \textit{Minecraft} there are approximately 1.38 trillion blocks per world \cite{III_2022}. Instead of relying on PCG for world creation, AI can analyze real-world photos from the internet and transform them into pixel-style 3D models. This allows for the generation of worlds that differ significantly from the basic natural terrain in \textit{Minecraft}, enabling the creation of environments like modern cities. Recent years some AI-powered generators such as \textit{BuilderGPT}\footnote{https://github.com/CubeGPT/BuilderGPT} developed by Zhou-Shilin have been adopted by users to assist to generate more creative maps, making the creation process automatic or semi-automatic. Moreover, Awiszus \etal \cite{awiszus2021world} proposed a generative model called ``World-GAN" for \textit{Minecraft} worlds, which is based on NLP and is applied to generating larger and more complex worlds while meeting the needs of specified terrain. Figure \ref{fig_disscussion_c} shows a comparison between traditional generation and AI-assisted generation.

While AI simplifies the most time-consuming part and reduces the cost of designing and creating new worlds for users, it is not entirely satisfactory at generating content that meets players' expectations, especially text-to-3D, which is almost unavailable \cite{shen2023effects}. However, the relevant technology is still evolving, and we believe that the future trend of UGC must be inseparable from the dialectical view of AIGC. As we mentioned in Section \ref{subsec_level editor}, the use of AI as a supplementary tool can not only lower the creation barriers but also enhance users' creative enthusiasm, since its role is particularly evident in processing large amounts of data or automating repetitive tasks. While the rise of AIGC may involve challenges such as copyright issues and cultural conflicts\cite{gervais2019machine}, we suggest that the potential benefits of AIGC outweigh these concerns if properly handled. It is crucial to deeply consider how to responsibly manage the application of AI in UGC creation.

\subsection{Ethical Consideration} \label{subsec_ethical}

\subsubsection{Cheating Phenomenon}

Multiplayer online games often raise ethical concerns regarding cheating, where cheaters employ software robot or manipulate execution code to gain an unfair advantage over other players \cite{laurens2007novel, yeung2006detecting}. Despite the implementation of strict anti-cheating measures, including detection systems and severe penalties to deter such behavior, the allure of potential fame and fortune continues to drive a significant number of users to cheat. Simultaneously, for game developers, the implementation of anti-cheating mechanisms and the enforcement of penalties for cheaters significantly increased the development and maintenance costs of the game \cite{blackburn2014cheating}, thereby exacerbating the conflict between players and the industries. In academia, Yan and Randell \cite{yan2005systematic} proposed a classification of cheating in online games in 2005, considering that fairness is a crucial factor influencing the prevalence of cheating. Consalvo \cite{consalvo2009cheating} defined cheating in both multiplayer and single-player games, which noted that many players believed using external tools to play a game after completing it wasn't cheating, though this view was generally confined to single-player experiences (which we defineas ``user-generated instruments" in Section \ref{subsec_instrument}). Broadly speaking, cheating involves the player's actions that break the game's rules. However, when discussing cheating, the primary concern is fairness in multiplayer games. In these contexts, cheaters gain a significant advantage over other players, disrupting the game's balance \cite{adams2014fundamentals}. On an ethical level, Szell and Thurner \cite{szell2010measuring} argued that the competitive and interactive nature of games mirrored real-world dynamics, suggesting that widespread cheating in games reflects a broader lack of moral awareness among players in society.

When it comes to UGC, if cheating is simply defined as breaking certain rules within the game, much of the UGC discussed earlier could be classified as cheating. Unlike third-party guides or tutorials that for instance help players obtain hidden achievements, cheating involves circumventing the game’s rules to gain an unfair advantage that other players cannot access. We notice that tools often considered cheating, such as auto-targeting modifiers, are also user-generated and a form of creativity \cite{hamlen2015problem}. Blackburn's research on Steam's user community \cite{blackburn2014cheating} suggests that though cheaters are largely indistinguishable from fair players, that users who are labelled as cheaters suffer a ``social penalty" of losing friends. Ethically, the primary harm of cheating lies in its disruption of game balance and the subsequent negative impact on the gaming experience, as well as the game's lifecycle. When discussing how to prevent the generation of cheat-related content, developing moral negative feedback mechanisms from a social and group psychological perspective may be a viable solution \cite{ellemers2012morality}.

\subsubsection{Copyright Infringement}

In the process of user generation, copyright infringements caused by weak copyright awareness are not rare. We categorize these conflicts into two main types. (1) The conflict between users and third-party copyright owners: The most obvious example is, when users (also noted as modders in some articles) introduce resource files from third-party products without permission, players who do not own those products can access the resources by downloading the relevant modification of the game. Additionally, when AI is employed as a tool in user generation, it often involves the use of unauthorized works during the model training process \cite{duan2024incentive}. On the other hand, there are someone believes that creators who make MODs for non-profit purposes should not have to pay a legal price for the freedom to create \cite{postigo2008video}. (2) The conflict between users and game developers: Some game industries, such as \textit{Atari}, have argued that games are their own property, and that any content created based on them is infringing \cite{hemnes1982adaptation}. While there is not yet a game that has completely rejected UGC, it is common for those game developers to limit the scope of what UGC can do, dividing the game into a part that users can change and a core that cannot be changed. Not only for UGC that changes the game itself, but also for media-based user-generated content such as game livestreams on \textit{YouTube} or related game-related videos are at risk of being sued by developers for infringement \cite{jungar2016streaming}.

Legally speaking, video games are more complex than other copyrighted products, while many local laws treat video games as computer programs \cite{abrams2013legal}. However, as video games have evolved, people have come to realize that the resource files (images and sounds) and gameplay in video games also need copyright protection, which is still controversial. For example, most FPS games on the market are basically the same in terms of gameplay. Copyright infringements are essentially conflicts of interest. Giving priority to the interests of the developer or copyright owner, while limiting the authorities of the user to create, stifles user's creativity and expression, and deprives the community of a way to benefit from the platform \cite{postigo2008video}. Regarding the solution of copyright problem in video games, we agree with Jungar's \cite{jungar2016streaming} view that the purpose of copyright is not to ``maximize revenue", but to ``incentivize innovation". Taking a negative attitude towards creators and completely denying the rationality of the existence of UGC are not conducive to the copyright owners to obtain more profits.

\subsection{Game, User and Community}\label{subsec_guc}

\begin{figure}[t!]
		\includegraphics[width=1.0\columnwidth]{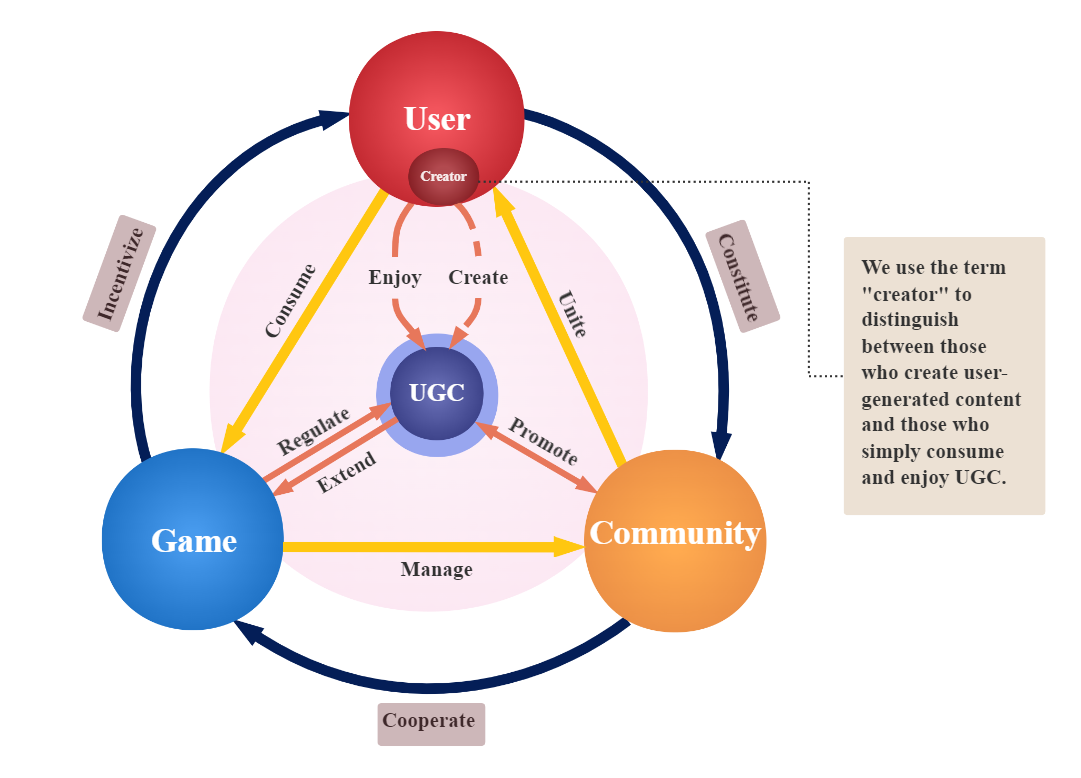}
	\caption{A graph of the relationship between video game, user, and community, including the connections between the three and user-generated content}
        \label{fig_disscussion_e}
\end{figure}

To envision the future of UGC in the video game industries, it is essential to understand the intricate relationship between video games, users, and communities \cite{postigo2007mods}. Figure \ref{fig_disscussion_e} presents a schematic diagram that visualizes the dynamic interaction between the three entities and highlights how UGC acts as a bridge, enhancing engagement and boosting creativity within video game industry. 

Sánchez \etal \cite{sanchez2012playability} analyzes the user experience in video games and puts forward the concept of playability, which proves that users consume video games. As an excellent test bed for Human Computer Interaction (HCI), users play a dominant role in the game design, which is called user-centered design \cite{ng2014review}. Since user is often referred to as the player or more generally the consumer, user experience is often the fundamental criterion for evaluating the quality of the game. Adam \cite{adams2014fundamentals} argues in his book that games have certain reward and punishment mechanisms that incentivizes sustained interest in the game. For example, in \textit{PUBG: Battlegrounds}\footnote{https://www.pubg.com/}, players who get killed cannot continue to participate in the competition, and the last player or team that survives will win the game. This is a typical reward and punishment mechanism, where the punishment does not mean that the player has broken the rules, but rather gives the player frustration through social comparison from a psychological point of view, similarly, the winners will reward positive emotions such as happiness \cite{garcia2013psychology}. Successful game designers use rewards and punishments to incentivize users and increase user engagement, thereby extending the lifecycle of the game.

Communities unite users around the world, while the social behaviors among users constitute communities. As a third party group that does not directly consume video games, community has often been regarded as a platform for users to share and communicate in past studies 
\cite{rheingold2000virtual}. Moreover, Hsu \cite{hsu2007consumer} recognizes the role of online communities in increasing customer loyalty and differentiates business communities from entertainment communities. Some communities like \textit{Planet Minecraft}\footnote{https://www.planetminecraft.com/} are created solely for specific games, while others, such as \textit{Nexus MODs}\footnote{https://www.nexusmods.com/}, exist as a popular platform for sharing MODs. With the development of network, the function of the community has gradually expanded to provide a platform for trading in-game items and other goods. The community usually extracts part of the profits from users' transaction to maintain the daily operation of the website. Besides, advertising, fees paid by a few privileged users, etc., also bring revenue to the community. In turn, the community inspires a sense of belonging among users, and the events and competitions held within the community increase user participation and provide emotional support to foster social ties \cite{taylor2009play}.

Game community is closely related to video games themselves \cite{burger2011user}, since community is one of the quickest ways to give developers timely feedback on game evaluation and user experience. While some game developers create and manage communities as spin-offs of video games, unofficial communities are not affiliated with any game. Even though most of those third-party communities claim to be non-profit, there are significant costs to running and maintaining the site, while, in fact, the community benefits from traffic and user attention. As we mentioned in \ref{subsec_ethical}, there is a potential conflict of interest between the community and the video game developer. However, when it comes to the relationship between video games and the community, we believe that official intervention to manage the community on a cooperative basis, where necessary, might be the best solution. For example, game developers have the right to intervene and police the community, where there is an interesting phenomenon within the communities that they encourage the exploitation of game bugs to gain an unfair competitive advantage.

The development of video games has benefited a lot from UGC \cite{lundmark2015effects}, and developers should have a positive attitude towards UGC considering about the prospect of UGC and video games, by regulating UGC appropriately, or integrating UGC into original games (e.g. \textit{Roblox}), and relying on the openness of UGC to drive higher monetization. In terms of content, UGC, as a derivative of video games, is intended to expand the vacancy not covered in the game. Daugherty’s research on consumer motivation \cite{daugherty2008exploring} suggests that UGC helps users alleviate self-doubt, while Xu \etal \cite{xu2017study} argues that UGC serves as a vital medium for players to express their emotions while gaming. In this study, we define ``creators" as users who possess creative abilities and actively engage in producing UGC. These creators have the ability to enjoy and consume UGC that all users have. Within the community, UGC plays a crucial role in fostering a sense of belonging and driving its long-term development of games \cite{sun2017motivation, kasapakis2017user}.

\section{Conclusion} \label{sec_conclusion}

In recent years, the development of video games and UGC has become increasingly intertwined, with UGC emerging as a crucial element in the future of the game industry. Based on previous studies, this paper refines the existing classification of in-game UGC and corresponding editors, addressing inconsistencies and ambiguities in earlier categorizations. In the discussion, we highlight the importance of UGC outside games that belongs to fandom content in game culture, which is not covered in the classification. We also speculate on the potential relation between game genres and the tendency of users to create UGC, providing insights into the impact of AI on user generation. In addition, ethical concerns surrounding cheating and copyright disputes in gaming are explored, with proposed solutions based on relevant literature. Finally, we examine the relationship between games, users, communities and UGC. We believe that the growth of UGC is positively correlated with the overall prospects of the video game industry, and the prosperity and innovation of UGC are poised to drive profitability within the game industry and academic research.

% {\appendix[Proof of the Zonklar Equations]
% Use $\backslash${\tt{appendix}} if you have a single appendix:
% Do not use $\backslash${\tt{section}} anymore after $\backslash${\tt{appendix}}, only $\backslash${\tt{section*}}.
% If you have multiple appendixes use $\backslash${\tt{appendices}} then use $\backslash${\tt{section}} to start each appendix.
% You must declare a $\backslash${\tt{section}} before using any $\backslash${\tt{subsection}} or using $\backslash${\tt{label}} ($\backslash${\tt{appendices}} by itself
%  starts a section numbered zero.)}

 

\bibliographystyle{IEEEtran}
\bibliography{IEEEabrv,reference}

\begin{thebibliography}{10}
\providecommand{\url}[1]{#1}
\csname url@samestyle\endcsname
\providecommand{\newblock}{\relax}
\providecommand{\bibinfo}[2]{#2}
\providecommand{\BIBentrySTDinterwordspacing}{\spaceskip=0pt\relax}
\providecommand{\BIBentryALTinterwordstretchfactor}{4}
\providecommand{\BIBentryALTinterwordspacing}{\spaceskip=\fontdimen2\font plus
\BIBentryALTinterwordstretchfactor\fontdimen3\font minus \fontdimen4\font\relax}
\providecommand{\BIBforeignlanguage}[2]{{%
\expandafter\ifx\csname l@#1\endcsname\relax
\typeout{** WARNING: IEEEtran.bst: No hyphenation pattern has been}%
\typeout{** loaded for the language `#1'. Using the pattern for}%
\typeout{** the default language instead.}%
\else
\language=\csname l@#1\endcsname
\fi
#2}}
\providecommand{\BIBdecl}{\relax}
\BIBdecl

\bibitem{min2019blockchain}
T.~Min, H.~Wang, Y.~Guo, and W.~Cai, ``Blockchain games: A survey,'' in \emph{2019 IEEE conference on games (CoG)}.\hskip 1em plus 0.5em minus 0.4em\relax IEEE, 2019, pp. 1--8.

\bibitem{naab2017studies}
T.~K. Naab and A.~Sehl, ``Studies of user-generated content: A systematic review,'' \emph{Journalism}, vol.~18, no.~10, pp. 1256--1273, 2017.

\bibitem{zhao2009survey}
Z.~Zhao, N.~Laga, and N.~Crespi, ``A survey of user generated service,'' in \emph{2009 IEEE International Conference on Network Infrastructure and Digital Content}.\hskip 1em plus 0.5em minus 0.4em\relax IEEE, 2009, pp. 241--246.

\bibitem{sun2017motivation}
Y.~Sun, X.~Dong, and S.~McIntyre, ``Motivation of user-generated content: Social connectedness moderates the effects of monetary rewards,'' \emph{Marketing Science}, vol.~36, no.~3, pp. 329--337, 2017.

\bibitem{kasapakis2017user}
V.~Kasapakis and D.~Gavalas, ``User-generated content in pervasive games,'' \emph{Computers in Entertainment (CIE)}, vol.~16, no.~1, pp. 1--23, 2017.

\bibitem{diaz2021building}
R.~L.~S. D{\'\i}az, G.~R. Ruiz, M.~Bouzouita, and K.~Coninx, ``Building blocks for creating enjoyable games—a systematic literature review,'' \emph{International Journal of Human-Computer Studies}, p. 102758, 2021.

\bibitem{estrella2017different}
A.~Estrella-Ram{\'o}n and F.~Ellis-Chadwick, ``Do different kinds of user-generated content in online brand communities really work?'' \emph{Online Information Review}, 2017.

\bibitem{christodoulides2012memo}
G.~Christodoulides, C.~Jevons, and J.~Bonhomme, ``Memo to marketers: Quantitative evidence for change—how user-generated content really affects brands,'' \emph{Journal of advertising research}, vol.~52, no.~1, p.~53, 2012.

\bibitem{ukpabi2018drives}
D.~C. Ukpabi and H.~Karjaluoto, ``What drives travelers' adoption of user-generated content? a literature review,'' \emph{Tourism management perspectives}, vol.~28, pp. 251--273, 2018.

\bibitem{barreda2013analysis}
A.~Barreda and A.~Bilgihan, ``An analysis of user-generated content for hotel experiences,'' \emph{Journal of Hospitality and Tourism Technology}, 2013.

\bibitem{cuomo2020user}
M.~T. Cuomo, D.~Tortora, A.~Giordano, G.~Festa, G.~Metallo, and E.~Martinelli, ``User-generated content in the era of digital well-being: A netnographic analysis in a healthcare marketing context,'' \emph{Psychology \& Marketing}, vol.~37, no.~4, pp. 578--587, 2020.

\bibitem{luca2015user}
M.~Luca, ``User-generated content and social media,'' in \emph{Handbook of media Economics}.\hskip 1em plus 0.5em minus 0.4em\relax Elsevier, 2015, vol.~1, pp. 563--592.

\bibitem{lastowka2007user}
G.~Lastowka, ``User-generated content and virtual worlds,'' \emph{Vand. J. Ent. \& Tech. L.}, vol.~10, p. 893, 2007.

\bibitem{lundmark2015effects}
J.~Lundmark and E.~Sandstr{\"o}m~Lindberg, ``The effects of intrinsic motivation, extrinsic motivation and toolkits onuser participation in user-generated content for video games:: A quantitative study of product development in online communities,'' 2015.

\bibitem{brunt2020influence}
C.~S. Brunt, A.~S. King, and J.~T. King, ``The influence of user-generated content on video game demand,'' \emph{Journal of Cultural Economics}, vol.~44, no.~1, pp. 35--56, 2020.

\bibitem{duan2022user}
H.~Duan, Y.~Huang, Y.~Zhao, Z.~Huang, and W.~Cai, ``User-generated content and editors in video games: Survey and vision,'' in \emph{2022 IEEE conference on games (CoG)}.\hskip 1em plus 0.5em minus 0.4em\relax IEEE, 2022, pp. 536--543.

\bibitem{daugherty2008exploring}
T.~Daugherty, M.~S. Eastin, and L.~Bright, ``Exploring consumer motivations for creating user-generated content,'' \emph{Journal of interactive advertising}, vol.~8, no.~2, pp. 16--25, 2008.

\bibitem{wirman2009productivity}
H.~Wirman, ``On productivity and game fandom,'' \emph{Transformative Works and Cultures}, vol.~3, no.~8, 2009.

\bibitem{sotamaa2015online}
O.~Sotamaa and H.~Wirman, ``Online games modifications and user generated content,'' \emph{The International Encyclopedia of Digital Communication and Society}, pp. 1--10, 2015.

\bibitem{postigo2008video}
H.~Postigo, ``Video game appropriation through modifications: Attitudes concerning intellectual property among modders and fans,'' \emph{Convergence}, vol.~14, no.~1, pp. 59--74, 2008.

\bibitem{liapis2013sentient}
A.~Liapis, G.~N. Yannakakis, and J.~Togelius, ``Sentient sketchbook: computer-assisted game level authoring,'' 2013.

\bibitem{ng2018situated}
G.~Ng, J.~G. Shin, A.~Plopski, C.~Sandor, and D.~Saakes, ``Situated game level editing in augmented reality,'' in \emph{Proceedings of the Twelfth International Conference on Tangible, Embedded, and Embodied Interaction}, 2018, pp. 409--418.

\bibitem{davidovici2009dynamics}
M.~Davidovici-Nora, ``The dynamics of co-creation in the video game industry: the case of world of warcraft,'' \emph{Communications \& Strategies}, no.~73, p.~43, 2009.

\bibitem{summerville2016super}
A.~Summerville and M.~Mateas, ``Super mario as a string: Platformer level generation via lstms,'' \emph{arXiv preprint arXiv:1603.00930}, 2016.

\bibitem{guzdial2016game}
M.~Guzdial and M.~Riedl, ``Game level generation from gameplay videos,'' in \emph{Proceedings of the AAAI Conference on Artificial Intelligence and Interactive Digital Entertainment}, vol.~12, no.~1, 2016, pp. 44--50.

\bibitem{awiszus2021world}
M.~Awiszus, F.~Schubert, and B.~Rosenhahn, ``World-gan: a generative model for minecraft worlds,'' in \emph{2021 IEEE Conference on Games (CoG)}.\hskip 1em plus 0.5em minus 0.4em\relax IEEE, 2021, pp. 1--8.

\bibitem{guzdial2019friend}
M.~Guzdial, N.~Liao, J.~Chen, S.-Y. Chen, S.~Shah, V.~Shah, J.~Reno, G.~Smith, and M.~O. Riedl, ``Friend, collaborator, student, manager: How design of an ai-driven game level editor affects creators,'' in \emph{Proceedings of the 2019 CHI conference on human factors in computing systems}, 2019, pp. 1--13.

\bibitem{quinlan1986induction}
J.~R. Quinlan, ``Induction of decision trees,'' \emph{Machine learning}, vol.~1, pp. 81--106, 1986.

\bibitem{postigo2007mods}
H.~Postigo, ``Of mods and modders: Chasing down the value of fan-based digital game modifications,'' \emph{Games and culture}, vol.~2, no.~4, pp. 300--313, 2007.

\bibitem{adams2014fundamentals}
E.~Adams, \emph{Fundamentals of game design}.\hskip 1em plus 0.5em minus 0.4em\relax Pearson Education, 2014.

\bibitem{Smith_Fodora_Guzik_2023}
\BIBentryALTinterwordspacing
R.~Smith, T.~Fodora, and I.~Guzik, ``Dota 2 vs. world of warcraft: What is the difference?'' Jan 2023. [Online]. Available: \url{https://www.buff.game/a/blog/dota-2-vs-world-of-warcraft-what-is-the-difference}
\BIBentrySTDinterwordspacing

\bibitem{KANG2024100697}
\BIBentryALTinterwordspacing
Y.~joo Kang, U.~jun Lee, and S.~Lee, ``Who makes popular content? information cues from content creators for users’ game choice: Focusing on user-created content platform “roblox”,'' \emph{Entertainment Computing}, vol.~50, p. 100697, 2024. [Online]. Available: \url{https://www.sciencedirect.com/science/article/pii/S187595212400065X}
\BIBentrySTDinterwordspacing

\bibitem{enwiki:1238577496}
{Wikipedia contributors}, ``Forza horizon 5 --- {Wikipedia}{,} the free encyclopedia,'' \url{https://en.wikipedia.org/w/index.php?title=Forza_Horizon_5&oldid=1238577496}, 2024, [Online; accessed 13-August-2024].

\bibitem{lin2022features}
Z.~Lin, H.~Duan, Z.~A. Wen, and W.~Cai, ``What features influence impact feel? a study of impact feedback in action games,'' in \emph{2022 IEEE Games, Entertainment, Media Conference (GEM)}.\hskip 1em plus 0.5em minus 0.4em\relax IEEE, 2022, pp. 1--6.

\bibitem{khalifa2016general}
A.~Khalifa, D.~Perez-Liebana, S.~M. Lucas, and J.~Togelius, ``General video game level generation,'' in \emph{Proceedings of the Genetic and Evolutionary Computation Conference 2016}, 2016, pp. 253--259.

\bibitem{Ipacs_2024}
\BIBentryALTinterwordspacing
D.~Ipacs, ``Level design: What is it and what makes it successful?'' Aug 2024. [Online]. Available: \url{https://bluebirdinternational.com/level-design/}
\BIBentrySTDinterwordspacing

\bibitem{summerville2018procedural}
A.~Summerville, S.~Snodgrass, M.~Guzdial, C.~Holmg{\aa}rd, A.~K. Hoover, A.~Isaksen, A.~Nealen, and J.~Togelius, ``Procedural content generation via machine learning (pcgml),'' \emph{IEEE Transactions on Games}, vol.~10, no.~3, pp. 257--270, 2018.

\bibitem{guzdial2018automated}
M.~Guzdial and M.~Riedl, ``Automated game design via conceptual expansion,'' in \emph{Proceedings of the AAAI Conference on Artificial Intelligence and Interactive Digital Entertainment}, vol.~14, no.~1, 2018, pp. 31--37.

\bibitem{cook2014ludus}
M.~Cook and S.~Colton, ``Ludus ex machina: Building a 3d game designer that competes alongside humans.'' in \emph{ICCC}, 2014, pp. 54--62.

\bibitem{laurens2007novel}
P.~Laurens, R.~F. Paige, P.~J. Brooke, and H.~Chivers, ``A novel approach to the detection of cheating in multiplayer online games,'' in \emph{12th IEEE International Conference on Engineering Complex Computer Systems (ICECCS 2007)}.\hskip 1em plus 0.5em minus 0.4em\relax IEEE, 2007, pp. 97--106.

\bibitem{ferreira2022fandom}
T.~M.~L. Ferreira, ``Fandom, culture, and videogames: Analysing what makes a successful videogame franchise through final fantasy,'' Ph.D. dissertation, Instituto Politecnico do Porto (Portugal), 2022.

\bibitem{jones2006shooting}
R.~Jones, ``From shooting monsters to shooting movies: Machinima and the transformative play of video game fan culture,'' \emph{Fan fiction and fan communities in the age of the Internet: New essays}, pp. 261--80, 2006.

\bibitem{li2010digital}
Q.~Li, ``Digital game building: Learning in a participatory culture,'' \emph{Educational research}, vol.~52, no.~4, pp. 427--443, 2010.

\bibitem{raessens2005computer}
J.~Raessens and J.~Goldstein, ``Computer games as participatory media culture,'' \emph{Handbook of computer game studies}, pp. 373--388, 2005.

\bibitem{dena2008emerging}
C.~Dena, ``Emerging participatory culture practices: Player-created tiers in alternate reality games,'' \emph{Convergence}, vol.~14, no.~1, pp. 41--57, 2008.

\bibitem{dym2018theyre}
B.~Dym, J.~Brubaker, and C.~Fiesler, ``theyre all trans sharon”: Authoring gender in video game fan fiction,'' \emph{Game Studies}, vol.~18, no.~3, pp. 1--28, 2018.

\bibitem{pearce2011communities}
C.~Pearce, \emph{Communities of play: Emergent cultures in multiplayer games and virtual worlds}.\hskip 1em plus 0.5em minus 0.4em\relax MIT press, 2011.

\bibitem{booth2015game}
P.~Booth, \emph{Game play: paratextuality in contemporary board games}.\hskip 1em plus 0.5em minus 0.4em\relax Bloomsbury Publishing USA, 2015.

\bibitem{gray2017fandom}
J.~Gray, C.~Sandvoss, and C.~L. Harrington, \emph{Fandom: Identities and communities in a mediated world}.\hskip 1em plus 0.5em minus 0.4em\relax NYU Press, 2017.

\bibitem{shen2023effects}
Z.~Shen, ``Effects of ai-generated content (aigc) in the game development: From traditional pcg to aigc,'' 2023.

\bibitem{yannakakis2018artificial}
G.~N. Yannakakis and J.~Togelius, \emph{Artificial intelligence and games}.\hskip 1em plus 0.5em minus 0.4em\relax Springer, 2018, vol.~2.

\bibitem{schaeffer2002games}
J.~Schaeffer and H.~J. Van~den Herik, ``Games, computers, and artificial intelligence,'' \emph{Artificial intelligence}, vol. 134, no. 1-2, pp. 1--7, 2002.

\bibitem{mi2023application}
X.~Mi, Y.~Cao, and Q.~Li, ``The application path of aigc in assisting game character design-a case study of the black bear monster from journey to the west,'' in \emph{2023 4th International Conference on Intelligent Design (ICID)}.\hskip 1em plus 0.5em minus 0.4em\relax IEEE, 2023, pp. 275--282.

\bibitem{bergensten2011short}
J.~Bergensten, ``A short demystification of the map seed,'' 2011.

\bibitem{III_2022}
\BIBentryALTinterwordspacing
R.~E.~W. III, ``How big is a minecraft world?'' Jan 2022. [Online]. Available: \url{https://www.lifewire.com/how-big-is-a-minecraft-world-5212822}
\BIBentrySTDinterwordspacing

\bibitem{gervais2019machine}
D.~J. Gervais, ``The machine as author,'' \emph{Iowa L. Rev.}, vol. 105, p. 2053, 2019.

\bibitem{yeung2006detecting}
S.~F. Yeung, J.~Lui, J.~Liu, and J.~Yan, ``Detecting cheaters for multiplayer games: theory, design and implementation,'' 2006.

\bibitem{blackburn2014cheating}
J.~Blackburn, N.~Kourtellis, J.~Skvoretz, M.~Ripeanu, and A.~Iamnitchi, ``Cheating in online games: A social network perspective,'' \emph{ACM Transactions on Internet Technology (TOIT)}, vol.~13, no.~3, pp. 1--25, 2014.

\bibitem{yan2005systematic}
J.~Yan and B.~Randell, ``A systematic classification of cheating in online games,'' in \emph{Proceedings of 4th ACM SIGCOMM workshop on Network and system support for games}, 2005, pp. 1--9.

\bibitem{consalvo2009cheating}
M.~Consalvo, \emph{Cheating: Gaining advantage in videogames}, 2009.

\bibitem{szell2010measuring}
M.~Szell and S.~Thurner, ``Measuring social dynamics in a massive multiplayer online game,'' \emph{Social networks}, vol.~32, no.~4, pp. 313--329, 2010.

\bibitem{hamlen2015problem}
K.~R. Hamlen and F.~C. Blumberg, ``Problem solving through “cheating” in video games,'' in \emph{Video games and creativity}.\hskip 1em plus 0.5em minus 0.4em\relax Elsevier, 2015, pp. 83--97.

\bibitem{ellemers2012morality}
N.~Ellemers and K.~van~den Bos, ``Morality in groups: On the social-regulatory functions of right and wrong,'' \emph{Social and Personality Psychology Compass}, vol.~6, no.~12, pp. 878--889, 2012.

\bibitem{duan2024incentive}
H.~Duan, A.~El~Saddik, and W.~Cai, ``Incentive mechanism design toward a win--win situation for generative art trainers and artists,'' \emph{IEEE Transactions on Computational Social Systems}, 2024.

\bibitem{hemnes1982adaptation}
T.~Hemnes, ``Adaptation of copyright law to video games,'' \emph{U. Pa. L. Rev.}, vol. 131, p. 171, 1982.

\bibitem{jungar2016streaming}
E.~E.~E. Jungar, ``Streaming video games: Copyright infringement or protected speech?'' \emph{Press Start}, vol.~3, no.~2, pp. 22--47, 2016.

\bibitem{abrams2013legal}
M.~S. Abrams, ``The legal status of video games: comparative analysis in national approaches,'' \emph{WIPO Magazine}, pp. 7--96, 2013.

\bibitem{sanchez2012playability}
J.~L.~G. S{\'a}nchez, F.~L.~G. Vela, F.~M. Simarro, and N.~Padilla-Zea, ``Playability: analysing user experience in video games,'' \emph{Behaviour \& Information Technology}, vol.~31, no.~10, pp. 1033--1054, 2012.

\bibitem{ng2014review}
Y.~Y. Ng and C.~W. Khong, ``A review of affective user-centered design for video games,'' in \emph{2014 3rd international conference on user science and engineering (i-user)}.\hskip 1em plus 0.5em minus 0.4em\relax IEEE, 2014, pp. 79--84.

\bibitem{garcia2013psychology}
S.~M. Garcia, A.~Tor, and T.~M. Schiff, ``The psychology of competition: A social comparison perspective,'' \emph{Perspectives on psychological science}, vol.~8, no.~6, pp. 634--650, 2013.

\bibitem{rheingold2000virtual}
H.~Rheingold, \emph{The virtual community, revised edition: Homesteading on the electronic frontier}.\hskip 1em plus 0.5em minus 0.4em\relax MIT press, 2000.

\bibitem{hsu2007consumer}
C.-L. Hsu and H.-P. Lu, ``Consumer behavior in online game communities: A motivational factor perspective,'' \emph{Computers in Human Behavior}, vol.~23, no.~3, pp. 1642--1659, 2007.

\bibitem{taylor2009play}
T.~L. Taylor, \emph{Play between worlds: Exploring online game culture}.\hskip 1em plus 0.5em minus 0.4em\relax MIT press, 2009.

\bibitem{burger2011user}
T.~Burger-Helmchen and P.~Cohendet, ``User communities and social software in the video game industry,'' \emph{Long Range Planning}, vol.~44, no. 5-6, pp. 317--343, 2011.

\bibitem{xu2017study}
Y.~Xu, D.~Li, F.~Si, and S.~Huang, ``Study on sentimental analysis for intermedia-ugc,'' in \emph{2017 2nd International Conference on Automation, Mechanical Control and Computational Engineering (AMCCE 2017)}.\hskip 1em plus 0.5em minus 0.4em\relax Atlantis Press, 2017, pp. 607--611.

\end{thebibliography}

\vspace{11pt}


\begin{IEEEbiography}[{\includegraphics[width=1in,height=1.25in,clip,keepaspectratio]{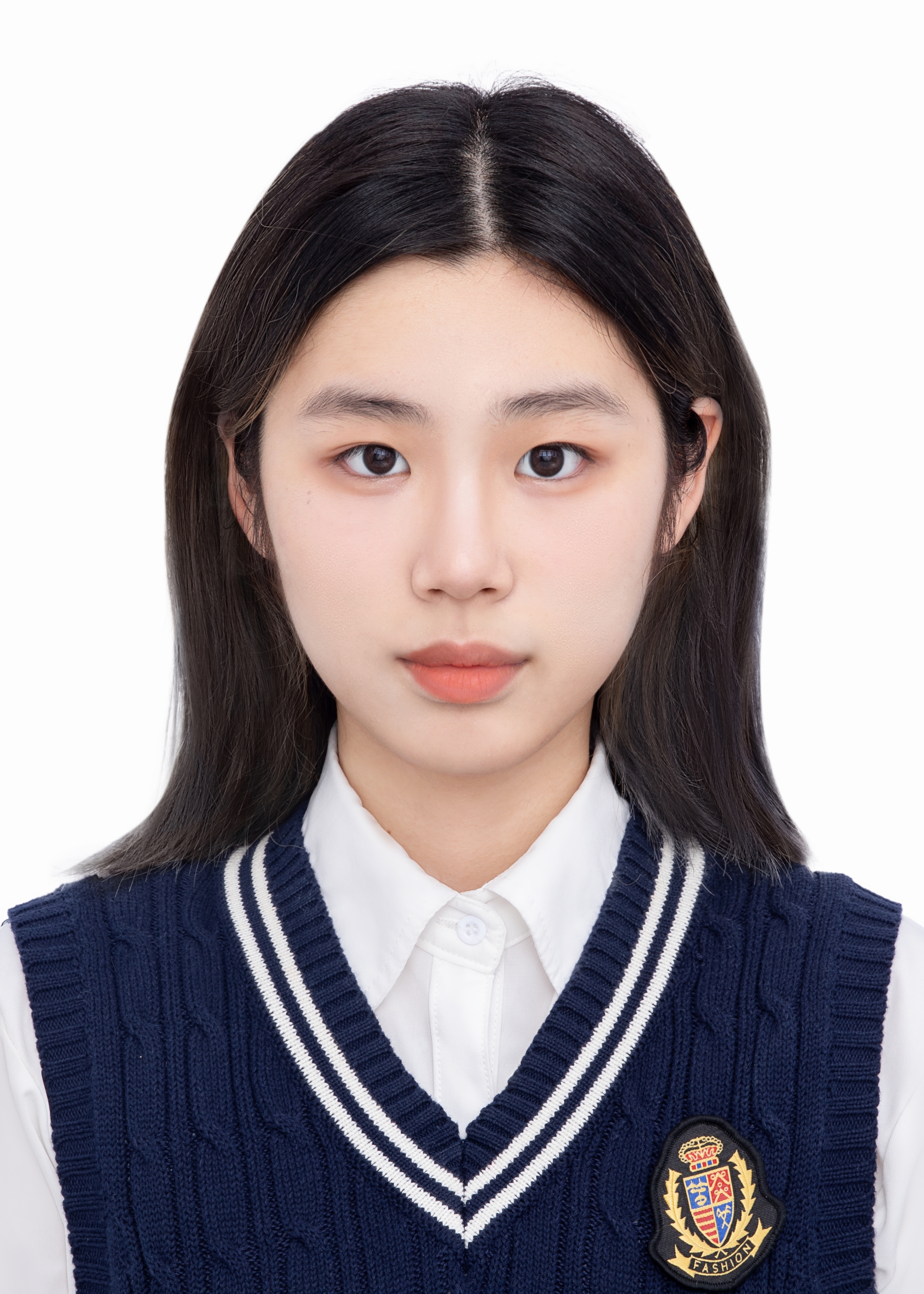}}]{Yuyue Liu} is currently pursuing her B.Eng. degree in Computer Science at School of Data Science, The Chinese University of Hong Kong, Shenzhen, China. She is also a visiting student at Guangdong-Hong Kong-Macao Joint Laboratory for Emotion Intelligence and Pervasive Computing, Artificial Intelligence Research Institute, Shenzhen MSU-BIT University, China. Her research interests include games, user-generated content.
\end{IEEEbiography}

\begin{IEEEbiography}[{\includegraphics[width=1in,height=1.25in,clip,keepaspectratio]{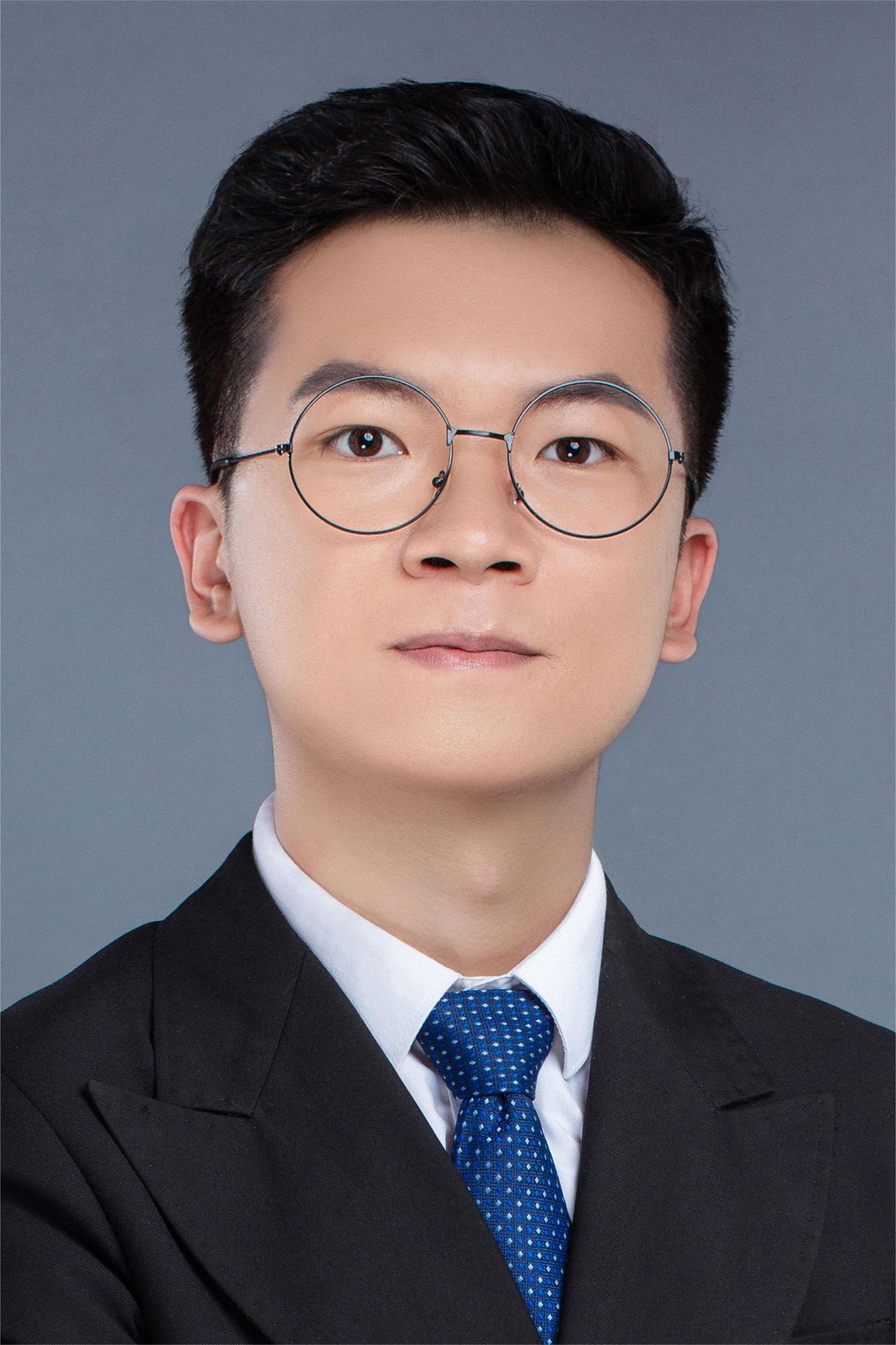}}]{Haihan Duan} (Member, IEEE) received his B.Eng. degree in Computer Science and Technology from East China Normal University, Shanghai, China, in 2017, and his M.Eng. degree in Software Engineering from Sichuan University, Chengdu, China, in 2020, and his Ph.D. degree in Computer and Information Engineering from The Chinese University of Hong Kong, Shenzhen, China, in 2023. He is currently an associate professor with Guangdong-Hong Kong-Macao Joint Laboratory for Emotion Intelligence and Pervasive Computing, Artificial Intelligence Research Institute, Shenzhen MSU-BIT University (SMBU). Before joining SMBU, he worked as a postdoctoral research fellow at University of Ottawa and Mohamed bin Zayed University of Artificial Intelligence (MBZUAI), located in Abu Dhabi, United Arab Emirates. His research interests include multimedia, blockchain and Web3, metaverse, human-centered computing, and medical image analysis. He has served as special session chair for IEEE MMSP'22, Conference Secretary of IEEE SmartIoT'24, and TPC member for top conferences including ACM MM, NOSSDAV, and IEEE ICME.
\end{IEEEbiography}

\begin{IEEEbiography}[{\includegraphics[width=1in,height=1.25in,clip,keepaspectratio]{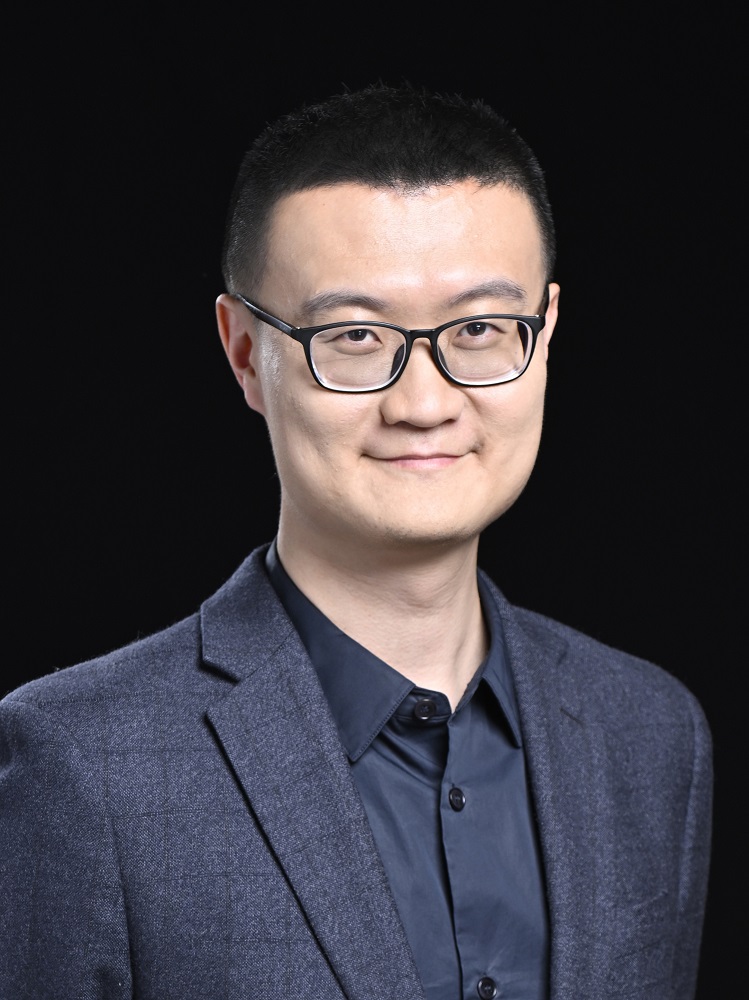}}]{Wei Cai} (Senior Member, IEEE) is a tenure-track Assistant Professor of Computer Science and Systems at the School of Engineering and Technology, University of Washington, Tacoma, WA, USA. He is now leading the Decentralized Computing Laboratory. He holds a Ph.D. in Electrical and Computer Engineering from The University of British Columbia (2016), an M.Sc. in Electrical Engineering and Computer Science from Seoul National University (2011), and a B.Eng. in Software Engineering from Xiamen University (2008). Prior to joining UW, Dr. Cai was an Assistant Professor of Electrical and Computer Engineering at The Chinese University of Hong Kong, Shenzhen, China. He has also conducted research visits at Academia Sinica (Taiwan), The Hong Kong Polytechnic University, and the National Institute of Informatics (Japan). Dr. Cai has co-authored over 100 peer-reviewed journal and conference papers and has received 6 Best Paper Awards. His research focuses on decentralized computing, with emphasis on mechanism design, social computing, multimedia, and applications. He serves as an Associate Editor for ACM Transactions on Multimedia Computing, Communications, and Applications (TOMM) and IEEE Transactions on Computational Social Systems (TCSS), and previously for IEEE Transactions on Cloud Computing (TCC). Dr. Cai is a Steering Committee member for ACM NOSSDAV, where he served as TPC co-chair in 2023, and has been an Area Chair for ACM MM since 2023. He is a Senior Member of IEEE and a member of ACM.
\end{IEEEbiography}

\vfill

\end{document}